\documentclass{aa}
\usepackage{hyperref}
\usepackage{graphicx}
\usepackage{natbib  }
\usepackage{color   }
\usepackage{array   }
\usepackage{txfonts }
\usepackage{breakurl}
\usepackage{multirow}

 \def\Jyb{\textrm{Jy~beam}$^{-1}$}       
 \def\kms{km~s$^{-1}$}                   
 \def\deg{$^\circ$}                      
 \def\am{$^\prime$}                      
 \def\as{$^{\prime\prime}$}              
 \def\hh{$^{\mathrm h}$}                 
 \def\mm{$^{\mathrm m}$}                 
 \newcommand{\cm}[1]{$\mathrm{cm}^{#1}$} 
 \newcommand{\dms}[3]{$#1^{#2}\!\!.#3$}  
 \def\CO{$^{13}\mathrm{CO}$}         
 \def\NH{$N_\mathrm{H}$}             
 \def\XMM{{\itshape XMM-Newton}}     
 \def\Chandra{{\itshape Chandra}}    

 \def\G29{G29.37+0.1}                

\begin{document} 

\title{Radio and X-ray properties of the source G29.37+0.1 linked to HESS~J1844$-$030}

\author{G. Castelletti \inst{1,2} 
   \and L. Supan \inst{1,2}
   \and A. Petriella \inst{2,3}
   \and E. Giacani \inst{2,4} 
\and B.C. Joshi \inst{5}}

\offprints{G. Castelletti}
\institute {Universidad de Buenos Aires, Facultad de Ciencias Exactas y Naturales. Buenos Aires, Argentina 
\and CONICET-Universidad de Buenos Aires, Instituto de Astronomía y Física del Espacio (IAFE), Buenos Aires, Argentina \\
\email{gcastell@iafe.uba.ar}
\and Universidad de Buenos Aires, Ciclo B\'asico Com\'un, Buenos Aires, Argentina 
               \and Universidad de Buenos Aires, Facultad de Arquitectura Dise\~no y Urbanismo, Buenos Aires, Argentina
            \and National Centre for Radio Astrophysics (NCRA), Post Bag No.3, Ganeshkhind, 411007, Pune, India }

\date{Received Month XX, 2016 / Accepted Month XX, 2016}

\abstract
 {}
{We report on the first detailed multiwavelength study of the radio source G29.37+0.1, which is 
an as-yet-unclassified object linked to the very-high-energy $\gamma$-emitting source HESS~J1844$-$030. 
The origin of the multiwavelength emission toward G29.37+0.1 has not been clarified so far, leaving open the question about the physical relationship
between these sources.}
{Using observations carried out with the Giant Metrewave Radio Telescope 
(GMRT), we performed high-quality full-synthesis imaging at 610~MHz of the field containing G29.37+0.1. The obtained data, combined with observations at 1400~MHz from The Multi-Array Galactic Plane Imaging Survey (MAGPIS) were used to 
investigate in detail the properties of its radio emission. Additionally, we 
reprocessed archival data obtained with the \it XMM-Newton \rm and \it Chandra \rm observatories in order to get a multiwavelength view of this unusual source.
 }
  {The radio source G29.37+0.1 mainly consists of a bright twisted structure, named the S-shaped feature. The high sensitivity of the new GMRT observations allowed the 
identification of potential lobes, jets, and a nuclear central region in the S-shaped morphology of G29.37+0.1.  
We also highlight the detection of diffuse and low surface brightness emission enveloping the brightest emitting regions. The brightest emission in G29.37+0.1 has a radio synchrotron spectral index $\alpha$=0.59$\pm$0.09. 
Variations in the spectral behaviour are observed across the whole radio source with the flattest spectral features in the central nuclear and jets components ($\alpha$$\sim$0.3). These results lead us to conclude
that the brightest radio emission from G29.37+0.1 likely represents a newly recognized radio galaxy.   
The identification of optical and infrared counterparts to the emission arising from the core of G29.37+0.1 strengthens our interpretation of an extragalactic origin of the radio emission. 
We performed several tests to explain the physical mechanism responsible for the observed X-ray emission, which appears overlapping the northeastern part of the radio emission. 
Our spectral analysis demonstrated that a non-thermal origin for the X-ray emission compatible 
with a pulsar wind nebula is quite possible. The analysis of the spatial distribution of the CO gas revealed the presence of a complex of molecular clouds located in projection adjacent to the radio halo emission 
and probably interacting with it. We propose that the faint halo represents a composite supernova remnant with
a pulsar powered component given by the diffuse X-ray emission superimposed along the line of sight to the 
radio galaxy. 
Further broadband observations of HESS J1844$-$030 are needed to disentangle 
its origin, although its shape and position suggest an extragalactic origin connected to G29.37+0.1.
}
  {}


\titlerunning{The puzzling G29.37+0.1/HESS~J1844$-$030 system}
\authorrunning{Castelletti et al.}

\maketitle

\section{Introduction}
Since their earlier classification as Fanaroff and Riley class I (FRI) and class II (FRII) objects \citep{fan74}, radio continuum surveys have been very efficient in identifying thousands of radio galaxies. While all these objects have a galactic nucleus, the main difference between the members of these two groups is the brightness surface distribution. Indeed, while FRI radio galaxies are core-brightened objects dominated by 
the presence of collimated jets extending in opposite directions from the energetic centre, the members of the FRII group also have hotspot regions of bright emission toward the end of diffuse lobes powered by the relativistic jets. Studies of radio galaxies highlight the
importance of these sources for understanding a variety of issues such as the cycle of radio activity and the interaction of the jets with the ambient medium during their evolution. 

While the radio galaxies are possibly the most energetic single entities 
in the Universe, supernovae (SNe) and their remnants (SNRs) could also be the most powerful astronomical sources inside a galaxy. In our Galaxy, despite considerable effort, the number of identified SNRs \citep[$\sim$300,][]{gre14} is at least three times smaller than that expected from statistical studies. The OB star counts, pulsar birth rates, and SN rates in other Local Group galaxies predict $\sim$1-2 stellar explosions per century  
and a lapse of 50,000 to 100,000~yr in which the radio synchrotron emission from 
their remnants is observable \citep{li91,tam94}.

Over the last decade, observations with instruments such as the \it Fermi\rm- Large Area Telescope (\it Fermi\rm-LAT), VERITAS, MAGIC, AGILE, and the High Energy Stereoscopic System (H.E.S.S.) have permitted successful identification of a huge amount of both Galactic and extragalactic 
$\gamma$-ray emitting sources. To date, a number of the high-energy emitting Galactic objects are known to be associated with SNRs, pulsar wind nebulae (PWNe), and binary systems, while active galactic nuclei (AGNs) would be responsible for the extragalactic $\gamma$-ray emission 
(\citealt{ace13,ack15,ace16}. For an extensive updated catalogue of TeV sources see, for instance, the TeVCat web page\footnote{\url{http://tevcat.uchicago.edu/}.}). There is still an important number of $\gamma$-ray sources, particularly in the Galactic plane, for which a clear identification has not been established. In these cases, a broadband approach using multiwavelength observations, especially in the radio and X-ray bands, is essential to advance the understanding of the nature of the $\gamma$-ray emission processes. 

This work focuses on the puzzling radio source named \object{G29.37+0.1}. 
The nature of this object has remained obscure: although G29.37+0.1 was proposed to be a new SNR candidate \citep{hel06}, its morphology comprising two oppositely-directed jets and lobes from a point source located at its centre strongly resembles the structure of a radio galaxy \citep[possibly PMN~J1844$-$0306\footnote{Parkes-MIT-NRAO (PMN) catalogue \citep{gre94}.},][]{hel89}. 
The situation became further confusing after the detection with the H.E.S.S. telescope array of a 
complex of TeV sources toward G29.37+0.1 \citep{hoppe08}.  
Among them, the so-called ``component~C'' (\object{HESS~J1843$-$033C}) was
detected overlapping only the northeastern edge of the radio source. The latest data recorded with H.E.S.S. reveal the $\gamma$-ray source \object{HESS~J1844$-$030} now spatially overlapping, at least as we see projected on the plane of the sky, the whole
radio emission from G29.37+0.1\footnote{Deil C. et al for the H.E.S.S. Collaboration, ``H.E.S.S. Galactic Plane Survey'', ICRC. August 4$^\mathrm{}$, 2015.}. 
As far as we know, no evidence for variability was reported in the gamma-ray flux. 
The preliminary analysis toward HESS~J1844$-$030 yields a flux of 1.0\% of the Crab's in the 0.2-100~TeV energy range. The apparent correlation makes a connection between the HESS source and G29.37+0.1 very plausible. Despite the intriguing morphology of G29.37+0.1 and its likely connection to HESS~J1844$-$030, this system has received little attention since it was first observed making this region worthwhile to survey. In this work, we present a multifrequency approach aimed at exploring the origin of both sources with a view to understanding whether they are physically linked or not. 

\section{Observations and data reduction}

\subsection{Radio continuum data}
We used the Giant Metrewave Radio Telescope (GMRT) during two sessions on 21 and 22 March 2015
to carry out full-synthesis imaging in the G29.37+0.1/HESS~J1844$-$030 region, with a total time of 17 hours under the project code 27$_{-}$027. The primary beam of the full array for imaging covers the entire radio emission from G29.37+0.1 (FWHM of the primary beam of GMRT is 0$^{\circ}$.7 at 610~MHz). 

The data from each day were fully reduced and imaged separately to ensure that there were no day-to-day amplitude discrepancies. All the calibration and data reduction were performed using the Astronomical Image Processing Software (AIPS) package developed by the N.R.A.O. We initiated the reduction of the data by manually removing  visibilities affected by strong radio frequency interference (RFI) throughout the observing band. Dead antennas were also flagged. To calibrate the amplitude response of individual antennas we used short observations of the standard calibrators 3C48 and 3C286 during March 21 and 22 observing runs, respectively. This calibration was performed setting the flux scale according to the coefficients taken from \citet{per13}. 
Phase and bandpass calibrations were derived from 1822$-$096. Subsequently, the calibration was 
applied to the target field visibilities. Before imaging, the data for each day were 
averaged in terms of frequency by collapsing the bandwidth to 41 spectral 
channels (the observing mode used a total BW 33~MHz comprising  512 individual 65~kHz channels). 
Such spectral average reduced the data volume and is acceptable to avoid radial smearing
out to the edge of the primary beam.

The final calibrated visibility data for
each day were combined into a single \it uv \rm  data set and imaged by adopting 
a weighting scheme with the Briggs robust parameter equal to 0 (a compromise between uniform weighting of the baseline for highest angular resolution and natural weighting for highest sensitivity). To deal with the non-coplanarity of the visibilities, we employed wide-field 
imaging based on a pseudo-three-dimensional multifacet algorithm \citep{cornwell92}.
In order to reduce residual phase variations and to increase the dynamic range, 
the imaging procedure also included two iterations of phase-only self-calibration and a final phase and amplitude self-calibration. Further self-calibration loops were not necessary as these did not improve the fidelity, nor reduce the noise of the image. In each iteration we inspected and flagged any weaker RFI-affected visibilities that could have persisted at this stage. The average synthesized beam for the map made with combined frequency channels is  
\dms{6}{\prime\prime}{55}~$\times$~\dms{5}{\prime\prime}{26}. The rms noise in the 610~MHz GMRT image is 0.42~mJy~beam$^{-1}$, after correcting for the response of the primary beam.

A complementary radio image at 1400~MHz available in the Multi-Array Galactic Plane Imaging Survey (MAGPIS, \citealt{hel06}) was also used to investigate the local spectral properties of G29.37+0.1 
in combination with the GMRT 610~MHz image. 
The MAGPIS image has a synthesized beam \dms{6}{\prime\prime}{2}~$\times$~\dms{5}{\prime\prime}{4} and an rms 0.7~mJy~beam$^{-1}$.

\subsection{X-ray observations}
\label{X-ray-observations}
We studied the X-ray emission from the {\G29} field by reprocessing archival data obtained
with the {\Chandra} and {\XMM} observatories. The combination of these two datasets permitted us 
to perform the most detailed spatially-resolved X-ray spectroscopic study of the region presented 
so far.

The source {\G29} was observed with {\Chandra} at two opportunities using the I array of the 
Advanced Charge-Coupled Device (CCD) Imaging Spectrometer (ACIS-I) operating in the VFAINT telemetry format. The region was also observed with {\XMM} in two opportunities using the three EPIC cameras aboard the telescope (i.e. MOS1, MOS2, and pn). Both observations were pointed toward the X-ray point-like source AX~J1845.0$-$0300, located in the northeastern portion of G29.37+0.1. 
This X-ray source was first discovered by \citet{tor98} with \it ASCA\rm. Table~\ref{table_X_1} summarizes the \it Chandra \rm and \it XMM-Newton \rm archival observations used in this paper.

The {\Chandra} data were reprocessed employing the {\Chandra} Interactive Analysis of Observations 
software package (CIAO, version 4.6) using the standard procedure in which the calibration products are provided by CALDB~4.6.1.1. The reduction of the {\XMM} observations was 
performed using the Science Analysis System (SAS, version 13.5.0) and the HEAsoft 
(version 6.15.1) analysis package. For each \it XMM-Newton \rm observation we obtained 
light curves in the high-energy band ($E$ > 10~keV) and we filtered the event files to remove periods of high count-rate. We applied additional filtering of the {\XMM} dataset to include only events with FLAG~=~0 and PATTERN~$\le12$ and $\le4$ for MOS and pn cameras, respectively. The final data consist of filtered event files appropriate to extract science products.

\begin{table*}
\caption{Summary of the {\Chandra} and {\XMM} observations in the region of {\G29}. We report the date of observation and the effective exposure time after filtering periods of high count-rate.}
\label{table_X_1}
\centering
\begin{tabular}{ccccc}                                                                                                           \hline\hline
 \multirow{2}{*}{Telescope} & \multirow{2}{*}{Obs-ID} & \multirow{2}{*}{Instrument} & \multirow{2}{*}{Date} & Effective exposure \\
                            &                         &                             &                       & time (ks)          \\\hline
 \multirow{2}{*}{\Chandra}  &           11232         &            ACIS-I           &      Aug.~11, 2009    &       29.5         \\
                              &           11801         &            ACIS-I           &      Jun.~17, 2010    &       29.4         \\\hline
 \multirow{2}{*}{\XMM}      &        0602350101       &        MOS1, MOS2, pn       &      Apr.~14, 2010    & 34.5, 34.5, 23.3   \\
                            &        0602350201       &        MOS1, MOS2, pn       &      Apr.~16, 2010    & 33.2, 37.5, 16.1   \\\hline                                   
\end{tabular}
\end{table*}

\subsection{The G29.37+0.1 field at other wavelengths}
For the purpose of deriving kinematic distances, we used the 21-cm line emission of the neutral hydrogen (HI) from the Very Large 
Array (VLA) Galactic Plane Survey (VGPS, \citealt{sti06}). 
These data imaged the region around G29.37+0.1 with a spatial resolution of 1$^{\prime}$, an rms noise level per velocity channel of $\sim$2~K, and a spectral resolution of 1.56~km~s$^{-1}$ with a separation in velocity channels of 0.82~km~s$^{-1}$. On the other hand, to investigate the morphological and kinematic properties of the molecular component of the interstellar medium (ISM), we employed data from the Galactic Ring Survey (GRS, \citealt{jac06}), which maps the J=1-0 rotational transition line of the $^{13}$CO gas with an angular resolution of 46$^{\prime\prime}$, a separation between consecutive velocity channels of $\sim$0.2~km~s$^{-1}$, and a sensitivity better than 0.4~K. 

\section{A multiwavelength view of the {\G29}/HESS~J1844$-$030 system}
\subsection{Radio continuum emission}
\subsubsection{Morphological properties}
\begin{figure*}[!ht]
  \centering
\includegraphics[width=16cm]{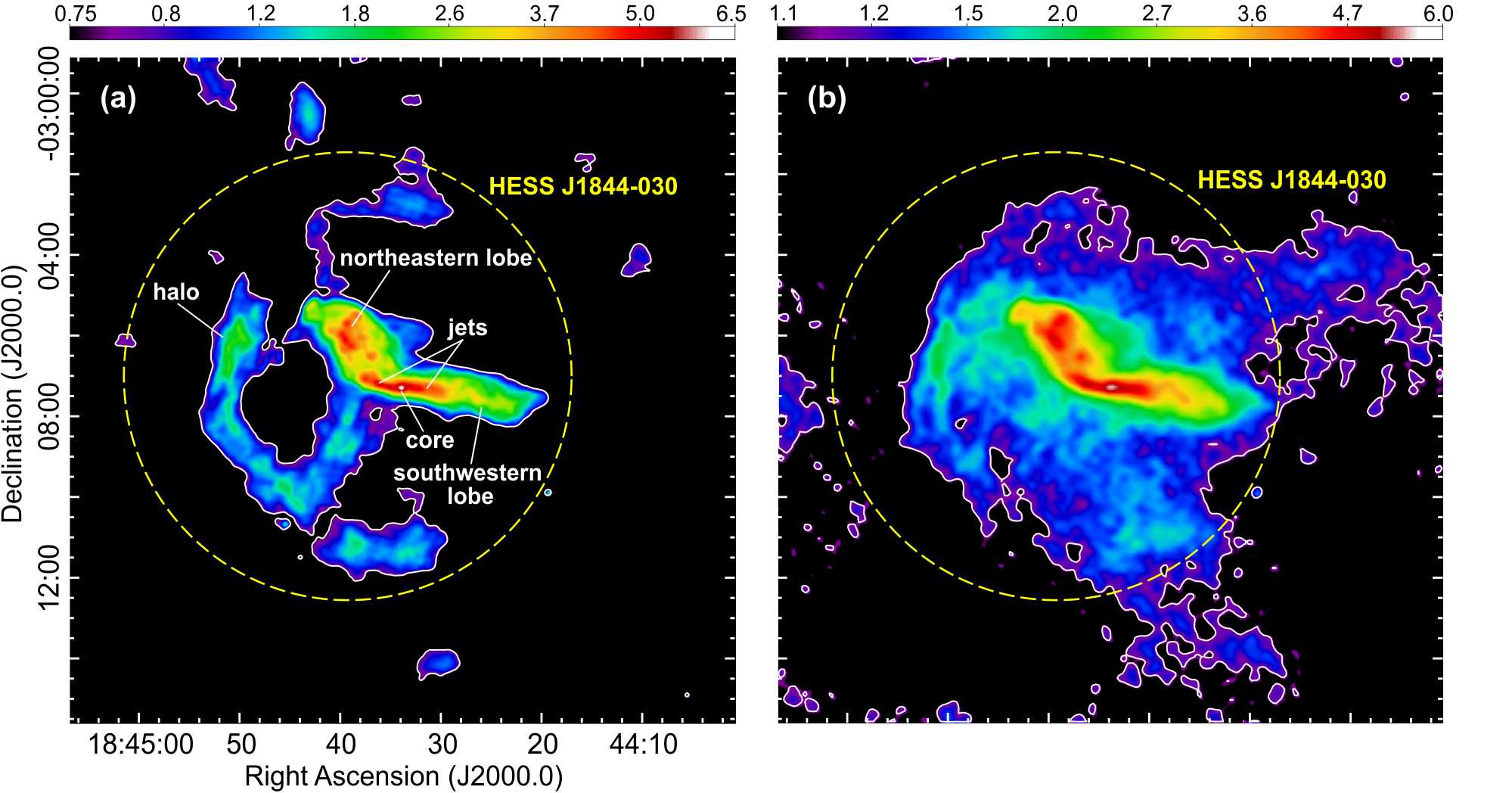}
\caption{\bf a) \rm Total power radio image at 610~MHz of the {\G29} field as observed with the GMRT.
Contour is traced at 1~{m\Jyb}.
The resolution is 
\dms{6}{\prime\prime}{55}~$\times$~\dms{5}{\prime\prime}{26} and the rms noise is $\sim$0.42~{m\Jyb};  
\bf b) \rm The same field of view at 1400~MHz from MAGPIS. Overlaid contour is at 
1.2~{m\Jyb}.
The angular resolution of this image is \dms{6}{\prime\prime}{20}~$\times$~\dms{5}{\prime\prime}{40} 
and the measured sensitivity level is $\sim$0.7~m\Jyb. The colour scales are both in {m\Jyb}. 
The dashed circle in both panels delineates the position and size of the TeV $\gamma$-ray source HESS~J1844$-$030.} 
  \label{radio-g29}
\end{figure*}

In Fig.~\ref{radio-g29}a we show the new GMRT 
total power continuum radio image of G29.37+0.1 at 610~MHz. The emission is characterized by a bright curved feature showing a ``S''-shaped 
profile extended about 6{\am} in size  in the northeast-southwest direction. As can be seen in the new GMRT image, this structure ends abruptly in the form of lobes with well defined outer boundaries. Significant internal curved structures are mainly noticeable in the 
northeastern portion of the arched emission, at the end of which a forward protrusion with enhanced radio emission is 
detected (RA$\simeq$18{\hh}44{\mm}43{\as}, Dec$\simeq$$-$03{\deg}05{\am}20{\as}, J2000.0). 
Inside the S-shaped feature, a double-sided jet seems to emanate from a central point-like source (hereafter, the ``nucleus'' or ``core'') located at 
RA$\simeq$18{\hh}44{\mm}34{\as}, Dec$\simeq$$-$03{\deg}07{\am}17{\as}. 
The projected separation between the centre of the source and the end of each jet is similar ($\sim$\dms{0}{\prime}{8}). Moreover, both radio lobes extend up to 
similar angular distances from the core 
($\sim$\dms{3}{\prime}{1} and $\sim$\dms{3}{\prime}{4} for the northeastern and southwestern lobe, respectively). A brightness asymmetry is observed between the lobes, being the northeastern one
nearly 2.7 times brighter than its opposite counterpart. 

\begin{figure}[!ht]
  \centering
\includegraphics[width=8cm]{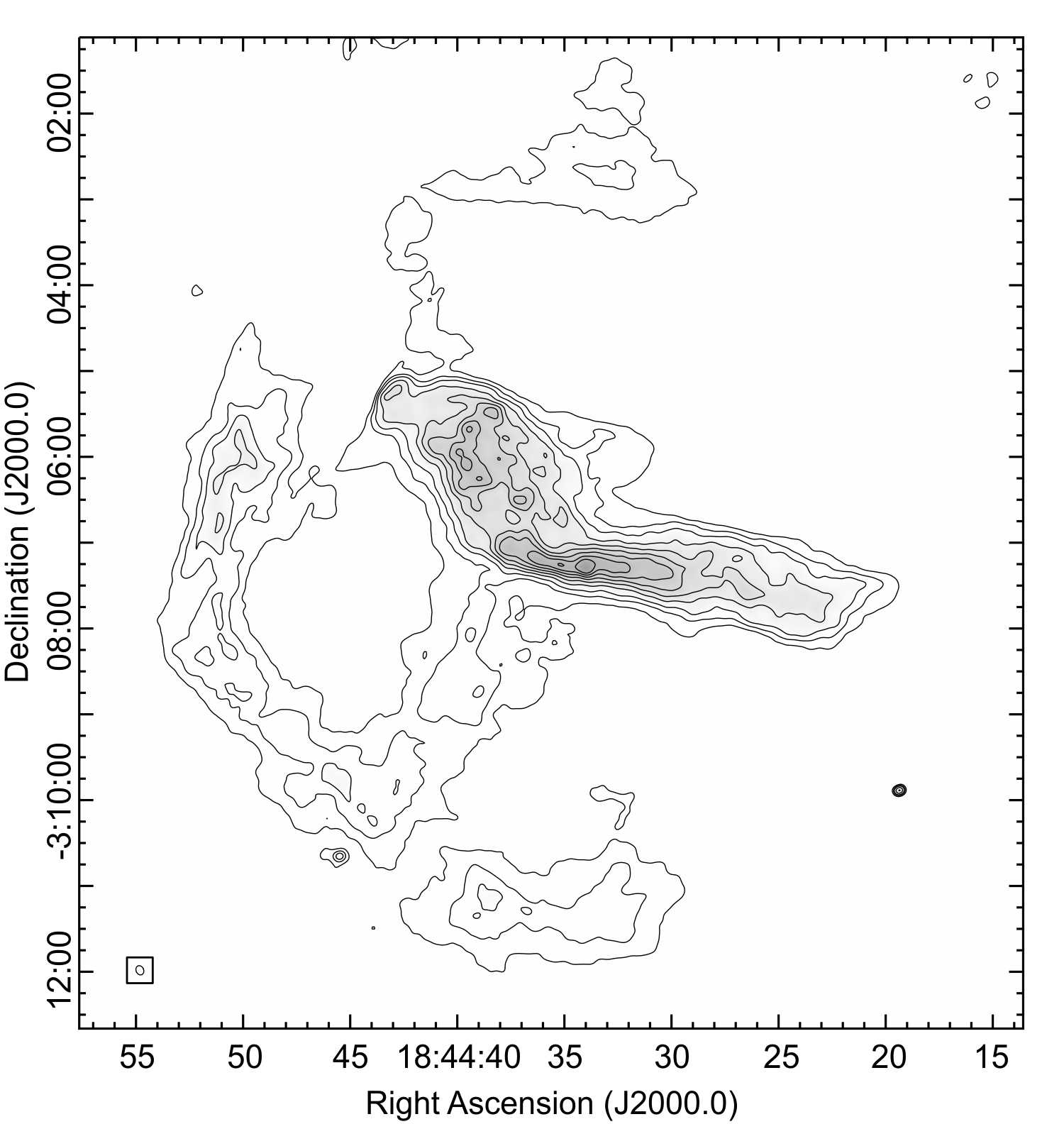}
  \caption{Radio contour map of G29.37+0.1 at 610~MHz. 
The contour levels traced at 0.8, 1.2, 1.6, 2.2, 2.9, 3.5, 4.1, 4.8, 
and 5.4~mJy~beam$^{-1}$ help to identify the presumed hotspot region and lobes 
in the S-shaped structure. The synthesized beam plotted at the bottom left corner is 
\dms{6}{\prime\prime}{55}~$\times$~\dms{5}{\prime\prime}{26}.}
  \label{contours}
\end{figure}

From a pure morphological point of view, the appearance of the S-shaped structure in
G29.37+0.1 is very reminiscent of a radio galaxy consisting of two opposite radio lobes 
at the end of a jet-nucleus structure \citep[e.g.][]{lar01,wez16}.
However, it is difficult to classify it as an FRI or FRII type galaxy 
by a simple morphological analysis. 
On one hand, G29.37+0.1 presents jets extending beyond a core, but these 
do not seem to dominate the radio emission, as is typically observed in FRI objects. 
On the other hand, edge-brightened lobe structures with bright radio emission inside 
are also part of the emission observed from this source, as is expected in galaxies belonging to the FRII category. 
Figure~\ref{contours} displays a total-intensity contour map of 
the full source at 610~MHz. 
In this representation the complex structure of brighter features are clearly shown. The bright emission region approximately at the middle 
of the northeastern lobe together with the protuberance at its outer edge may be interpreted as evidence for the so-called hotspot-like characteristic of FRII radio sources. Such protrusions of emission pushing forward the leading edge 
of the lobes were also observed in several radio galaxies classified as FRII sources 
\citep[see for instance 3C~223 and 4C~73.08,][]{wez16,orr10}. On the opposite southwestern lobe, the emission appears more diffuse without a significant hotspot.
Radio structural asymmetries, such as those observed in G29.37+0.1, were also found in 
other radio galaxies, J1211+743 and J1918+742 being good examples 
\citep[][and references therein]{pirya11}. 
Both intrinsic differences between the lobes and extrinsic effects (including orientation and environmental asymmetries) have been used by a number of authors to explain 
the observed asymmetries in the brightness of the lobes in radio galaxies \citep{gop00,pirya11}. 

The S-shaped feature in G29.37+0.1 is surrounded by a region of low surface brightness, which we designate ``halo''.
This faint emission has a relatively sharp outer boundary and is slightly brighter over the eastern and southern sides. We note that although artefacts are not present in the new GMRT image, 
the non detection of the complete emission from the halo at 610~MHz could be ascribed to possible missing low-level extended emission. At this frequency the halo component looks like a ring rather than an envelope of low-surface brightness 
distribution as is observed at 1400~MHz.
We also mention that this faint component is either undetected or poorly detected in other 
imaging radio surveys (e.g. 1.4~GHz NRAO VLA Sky Survey, \citealt{con98}; 1.4~GHz 
VGPS, \citealt{sti06}; NRAO VLA Archive Survey image at 5~GHz, \citealt{cros08}). 
This structure has probably been missed due to sensitivity limits in these radio observations plus the poorer overall \it uv\rm-coverage of snapshot in some of these surveys. 
The detection of similar halo components in radio galaxies is rare, with only few clear cases known. 
The most extreme example of a 
low-surface brightness component is M87 for which \citet{owen16}, using the Very Large Array 
at 330~MHz, reported the detection of a large-scale radio structure 
encompassing the lobes, jets, and nucleus of the radio galaxy. The faint emission in this region is 
inhomogeneous and highly filamentary.
Recently, \citet{wez16} after combining interferometric and single dish observations at 1400~MHz 
were able to resolve out the diffuse radio emission forming the halo structure in five radio galaxies (over a sample of 15 objects). 
Notably, all of them correspond to objects located well off the Galactic plane. 
In our work, the faint emission surrounding the S-shape feature in G29.37+0.1 could be either 
the halo of the radio galaxy or a Galactic feature in positional coincidence 
along the line of sight toward the S-like structure. In Sec.~\ref{CO} we discuss further this topic.

\subsubsection{Global and local radio spectral properties of G29.37+0.1}
\label{radio-spectrum}
Based on the new radio observations at 610~MHz, we have estimated an integrated flux density
of S$_{\mathrm{610}}$=1.42$\pm$0.35~Jy for the S-like feature in 
G29.37+0.1, where the attenuation of the primary
beam has been corrected for this calculation. In Table~\ref{S-fluxes} we list the 
estimated values of the flux densities integrated over the entire bright 
S-like component by using the new and public-domain data. The  corresponding flux 
measurements over its northeastern and southwestern radio lobes are also reported in 
Table~\ref{S-fluxes}. In all cases, the relative contribution of the nucleus and jet 
components to the total radio flux density of the whole S-shaped feature 
is slightly less than 10\%.
At each frequency, the area of integration was defined using a polygon around the emission region. 
All the new measured flux estimates were tied to the 
flux scale of \citet{per16}. 
The quoted errors are mainly due to uncertainties in defining the boundaries of the radio emission and the average background level that may occasionally be present, with
a contribution from the uncertainty in the calibration of each dataset.

\begin{figure}[!ht]
  \centering
\includegraphics[width=8cm]{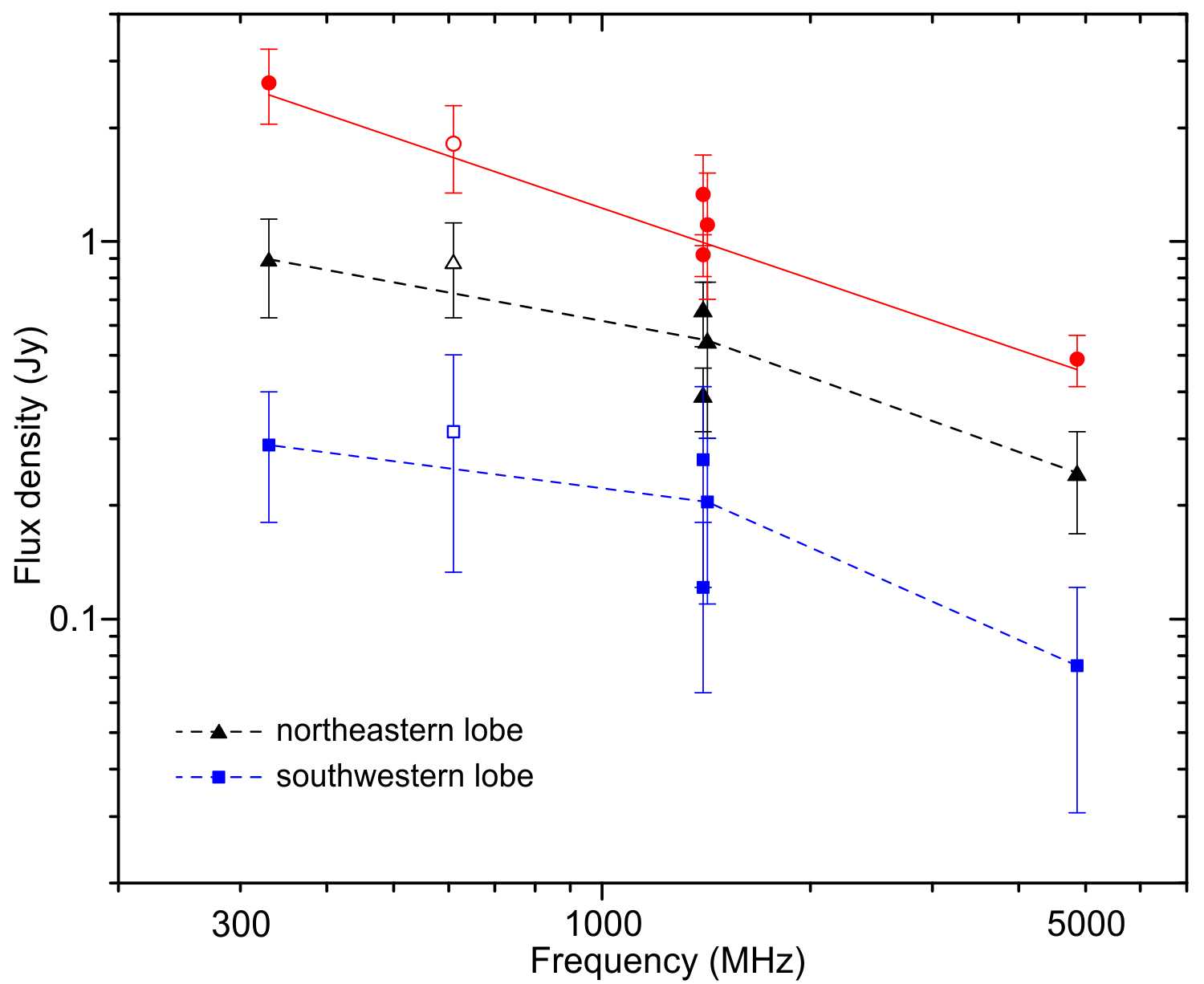}
  \caption{Total integrated spectrum for the S-shaped component and each radio
lobe in G29.37+0.1. The flux density measurements made using the new GMRT image 
are indicated by an open symbol. The straight line represents a weighted fit with 
a power law index $\alpha$=0.59$\pm$0.09 to the data points measured in the 
S-like feature.
No fit was performed to the flux density estimates in both radio lobes (the dashed line
between points is only a reference to the reader).}
  \label{spectrum}
\end{figure}

In Fig.~\ref{spectrum} we show a plot of the total integrated spectrum for 
the S-shaped feature and the two radio lobes in G29.37+0.1, calculated using
the flux densities reported in Table~\ref{S-fluxes}. 
A weighted fit to the data points using a single power law yields for the whole S-shaped
component a slope $\alpha$=0.59$\pm$0.09 (defined here by the relation 
$S_{\nu} \propto \nu^{-\alpha}$, with $S_{\nu}$ the flux at the frequency $\nu$). 
This integrated  spectral index implies 
that the radiation from the S-shaped feature has a non-thermal origin. 
The fact that the integrated flux density measured at 610~MHz falls 
nicely on the power law demonstrates that our GMRT image is missing little, if any emission on the angular scales of the lobes, and thus our flux density measurements and spectrum 
over the brightest S-like component are reliable. 

From Fig.~\ref{spectrum} and Table~\ref{S-fluxes} it seems that the ratio between the integrated flux densities of the lobes has a break at a frequency of $\sim$1400~MHz 
(a behaviour that is not observed in the integrated spectrum of the whole S-like feature).
The break is more noticeable for the southwestern lobe, which might be interpreted as a
physical difference between the two lobes (see Fig.~\ref{spectrum}). 
The departure from a single power-law could be attributed to processes of energy injection and losses that can occur as the energy carried by the jets is diffused into them \citep{pacho70}. 
A similar spectral behaviour showing a high frequency steepening was, 
for example, also observed in the integrated spectra of the radio lobes in the radio 
galaxy 4C~43.15 \citep{mor16}. 
In our case, the interpretation of a spectrum with a break for the radio lobes 
relies mainly on the flux density points at 330 and 4860~MHz. So, further observations toward G29.37+0.1, especially at low and high radio frequencies, are crucial to confirm the observed spectral trend. 
 
\begin{table}[h!]
\caption{Measured radio flux densities over the S-shaped feature and each lobe in G29.37+0.1.}
 \small
 \setlength{\extrarowheight}{5pt}
 \begin{tabular}{ccccc}
\hline\hline
   Frequency & \multicolumn{3}{c}{Flux density [Jy]} & Data \\ 
    \multicolumn{1}{c}{[MHz]}     &  S-shape & NE\tablefootmark{a} Lobe & SW\tablefootmark{a} Lobe & \\  \hline
       330        &  2.02$\pm$0.43  & 0.72$\pm$0.20 & 0.25$\pm$0.09 & GPS \\ 
       610        &  1.42$\pm$0.35  & 0.71$\pm$0.19 & 0.27$\pm$0.15   & This work \\
      1400        &  1.06$\pm$0.27  & 0.54$\pm$0.10 & 0.23$\pm$0.12 & MAGPIS\\
      1400        &  0.75$\pm$0.09  & 0.33$\pm$0.06 & 0.11$\pm$0.05 & NVSS \\
      1420        &  0.89$\pm$0.31  & 0.45$\pm$0.19 & 0.18$\pm$0.08 & VGPS\\
      4860        &  0.41$\pm$0.06  & 0.21$\pm$0.06 & 0.07$\pm$0.04 & NVAS\\\hline
\label{S-fluxes}
 \end{tabular}
\tablefoot{
\tablefoottext{a}{Here, NE and SW respectively denote the northeastern and southwestern radio lobes.}
\tablebib{GPS: VLA Galactic Plane Survey;
NVSS: NRAO VLA Sky Survey, \citet{con98}; 
MAGPIS: Multi-Array Galactic Plane Imaging Survey, \citet{hel06}; 
VGPS: VLA Galactic Plane Survey, \citet{sti06};
NVAS: NRAO VLA Archive Survey can be browsed through http://archive.nrao.edu/nvas/
}
}
\end{table}

\begin{figure}[!ht]
\centering
\includegraphics[width=8cm]{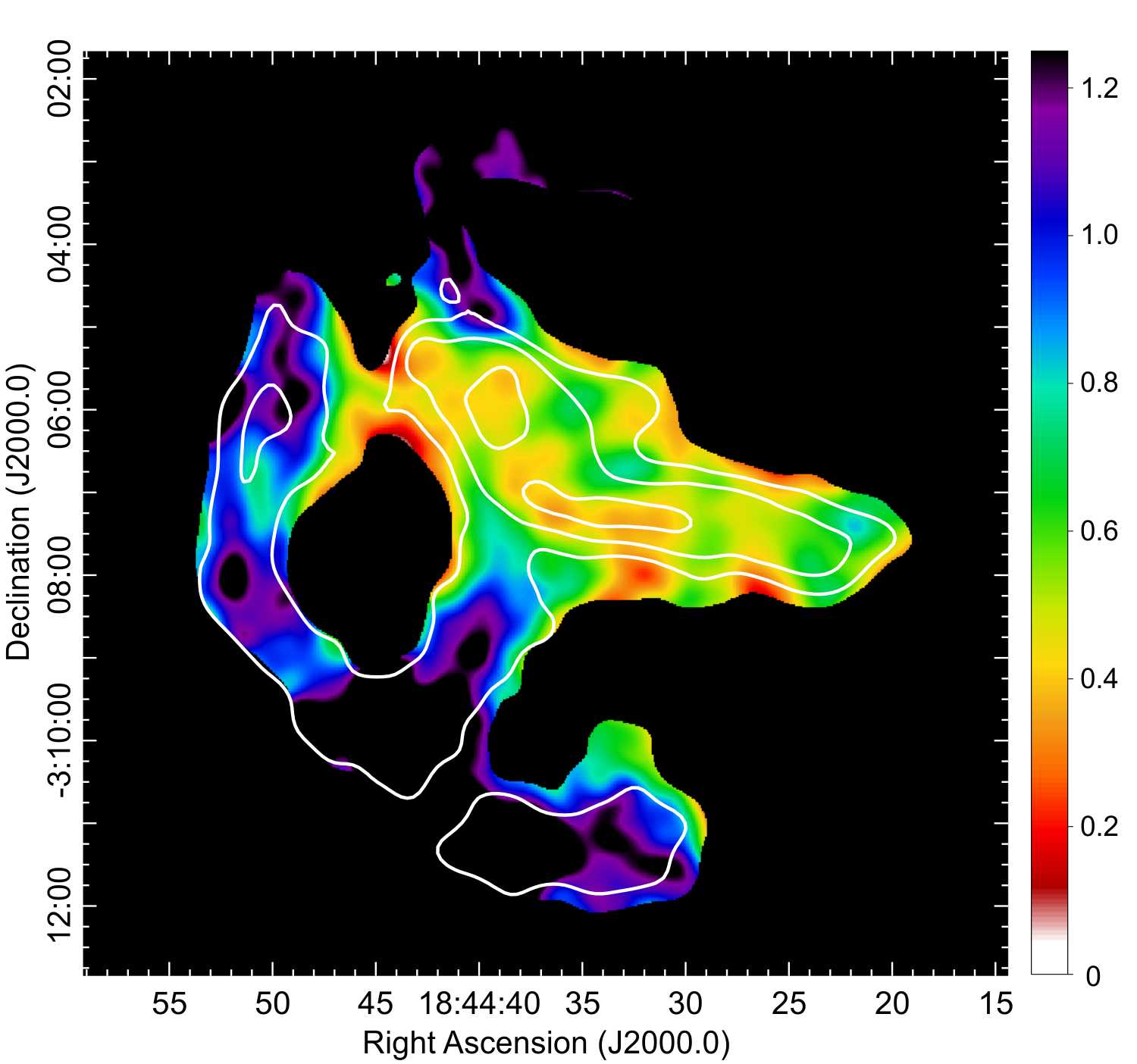}
\caption{Spatial spectral index distribution in G29.37+0.1 
in the interval of frequencies 610-1400~MHz performed using the new GMRT and the MAGPIS images (the spectral index follows the definition 
$S_{\nu}$$\propto$$\nu^{-\alpha}$). 
Only regions with flux densities greater than 4.3$\sigma$ and 3$\sigma$ of the respective noise levels at 610 and 1400~MHz 
were used to create the spectral index map. The 10, 20, and 40~mJy~beam$^{-1}$ contour levels from the 20$^{\prime\prime}$ resolution 
610~MHz image is included.}
\label{spectralmap}
\end{figure}

In addition to the global radio continuum spectra, we searched for spectral changes 
over the different components of the radio source G29.37+0.1. Local variations in the spectral radio index might be interpreted in terms of differences in the underlying physical process that  determine the energy of the particles. 
For this analysis we used the new GMRT image at 610~MHz, together with the image at 1400~MHz from MAGPIS. The latter image combines data from the B, C, and D configurations of the VLA with data from the Effelsberg 100~m telescope and thus does not suffer from the usual interferometric loss in flux. We cautiously note that the datasets from which the spectral indices were derived are not ideally matched to the same \it uv \rm range, as the visibility data were not available for the 
image at the higher frequency. 
We hence constructed the spatially resolved spectral index 
map shown in Fig.~\ref{spectralmap} by convolving 
both frequency images to a common synthesized beam 
of 20$^{\prime\prime}$, which abstains from any masking effect caused by small scale variations. 
For the S-like feature in G29.37+0.1 this approach 
is justified since the reconstruction of the largest
and smallest structures is similar at the two frequencies. 
This situation, however, does not hold for the halo component whose 
low surface brightness, together with some missing flux in the 610~MHz interferometric map, 
precludes any detailed spectral analysis. 
To avoid any positional offsets, the two images were aligned and interpolated to identical projections before calculating spectral indices. 
In addition, to create the spectral map
the 610 and 1400~MHz images were masked at the 4.3$\sigma$ and 3$\sigma$ significance level of their respective noise levels. 
The uncertainties in the determination of the spectral index from the map range between 0.08 and 0.15 for the brightest  and the faintest
regions of G29.37+0.1, respectively.  

The spectral index map displayed in Fig.~\ref{spectralmap} shows that the nucleus and the jet region in the S-shaped feature have a flat spectrum with a mean value 
$\alpha^{1400}_{610}$$\sim$0.3, while the northeastern and southwestern radio lobes have average
spectral indices of $\alpha^{1400}_{610}$$\sim$0.47 and $\alpha^{1400}_{610}$$\sim$0.55, respectively. 
In the northeastern side there is a region 
which has a flatter spectrum of about 0.38-0.45 
(compared with the surrounding lower brightness region of the lobe for which the spectrum ranges from 0.55 to 0.7). 
This flat spectral region coincides with the bright area observed inside this lobe, 
which we tentatively associated with a hotspot-like structure (see Fig.~\ref{contours}).
Other interesting spectral behaviour occurs in the north-easternmost region of G29.37+0.1, 
where a very flat spectrum ($\alpha^{1400}_{610}$$\sim$0.35-0.40) at the site where the X-ray emission was detected (as we discuss later in Sec.~\ref{secc_imag_X}). We noticed that the spectral behaviour in the S-like feature is in concordance with that determined in others FRII objects (see for instance \citealt{jam08} or \citealt{orr10}) and in radio sources with an hybrid appearance sharing properties of the FRI and FRII-type galaxies \citep{pirya11}. 

\subsection{X-ray analysis of G29.37+0.1}
\label{section-X}
\subsubsection{X-ray imaging}
\label{secc_imag_X}
To analyze the X-ray surface brightness in G29.37+0.1, we constructed images in different energy bands using the \it Chandra \rm and \it XMM-Newton \rm observations. 
The \it Chandra \rm data cover the overall radio emission from G29.37+0.1 including the halo region, while the \it XMM-Newton \rm field completely covers the jet and radio lobes but partially covers the halo seen at radio frequencies.
The morphology in the X-ray band as seen by both satellites is
similar and, due to its higher angular resolution, \it Chandra\rm's images reveal some interesting features that will be pointed out in the next paragraphs. 

\begin{figure}[ht]
  \centering
\includegraphics[scale=0.75]{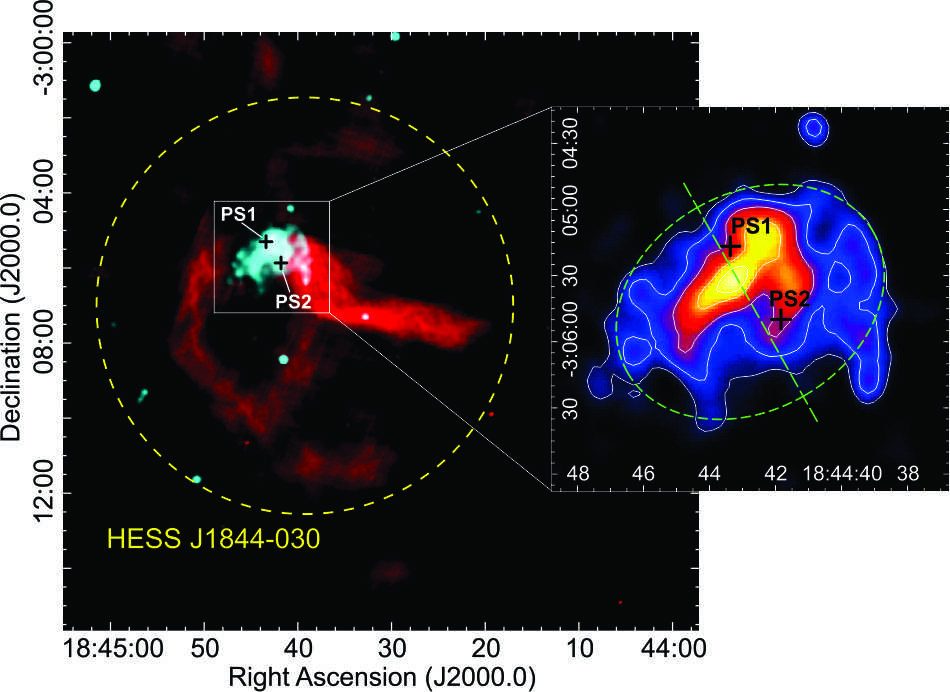}
  \caption{Colour image comparing the X-rays from {\G29} detected with \it Chandra \rm between 2.5 and 7.5~keV (Obs-ID 11232 and 11801) (in cyan) with radio continuum emission at 610~MHz (in red). The locations of the two point sources detected in X rays are marked using plus symbols. The circle indicates the size and position for the TeV emission from HESS~J1844$-$030. The inset depicts the X-ray emission after removing the point sources PS1 and 
PS2.
The ellipse in the inset indicates the region used to extract the spectrum of the entire nebula (see Sec.~\ref{secc_spec_X}). The dashed line marks the major-axis of the ellipse.}
  \label{fig_X1}
\end{figure}

Using the CIAO tool {\ttfamily merge$_{-}$obs}, we combined observations from the projects 11232 and 11801 to construct a mosaicked image of the region. 
The X-ray image was constructed with a spatial binning of three pixels, which results in a spatial scale of $\sim$\dms{1}{\prime\prime}{5}/pixel. 
The resulting image was then smoothed by using a Gaussian kernel with a radius of 2{\as}.
Figure~\ref{fig_X1} displays the \it Chandra \rm image obtained in the energy range 
2.5-7.5~keV, in which the X-ray morphology is better distinguished. 
As can be easily observed, the X rays coincide with the northeastern radio lobe of G29.37+0.1. 

Two compact sources were identified in both \it Chandra \rm and \it XMM-Newton \rm datasets. 
For ease of reference, we will call these sources PS1 and PS2 
(see Fig.~\ref{fig_X1}). 
The X-ray emission from PS1 originates mainly in the hard energy band (similar to the diffuse emission), while the emission from PS2 comes also from the soft and medium energy bands (i.e. 0.7-7~keV). We determined the centre of both point sources in the full resolution (i.e. spatial binning = 1) {\Chandra} mosaicked image by running the tool {\ttfamily wavdetect} in CIAO using the default settings. This tool requires a map of the point spread function (PSF), which was obtained by averaging the individual PSF maps of each observation, weighted by the respective exposure time.\footnote{Refer to the CIAO thread \url{http://cxc.harvard.edu/ciao/threads/}.} 
To highlight the diffuse emission, we removed the detected discrete X-ray sources from the image before adaptively smoothing it. We used the {\ttfamily roi} tool in CIAO to create the source and background regions for PS1 and PS2, taking into account the output of the {\ttfamily wavdetect} tool. Finally, we replaced the source region with an estimate of the local background using {\ttfamily dmfilth} with POISSON statistic.

A zoomed view of the diffuse X-ray emission in the field (after removing PS1 and PS2) 
is shown in the inset of  Fig.~\ref{fig_X1}. 
The asymmetric morphology of the X rays can be enclosed in an elliptical shape with an 
angular size of 65{\as}~$\times$~41{\as} on the plane of the sky and its semi-major axis oriented in the northeast-southwest direction.
In addition, the close-up image unveils two faint tongues of X-ray emission emanating from the southeast ($\sim$\dms{27}{\prime\prime}{5} long and $\sim$16{\as} thick) and the southwest ($\sim$40{\as} long and $\sim$22{\as} thick) sides of the nebula. These are oriented essentially
in the north-south direction. 

The {\Chandra} observations provide a sufficient number of photons to determine the position of PS1 and PS2 very accurately and hence
to identify potential optical and/or infrared counterparts to these sources. 
From the output of the {\ttfamily wavdetect} tool, 
for the source PS1 we obtain an elliptical source region of \dms{3}{\prime\prime}{99}~$\times$~\dms{2}{\prime\prime}{84} centred at RA=18{\hh}44{\mm}\dms{43}{\mathrm{s}}{369}, 
Dec=$-$03{\deg}05{\am}\dms{18}{\prime\prime}{19}.
The central position for the source PS2, derived inside a region \dms{1}{\prime\prime}{50}~$\times$~\dms{1}{\prime\prime}{35}, is RA=18{\hh}44{\mm}\dms{41}{\mathrm{s}}{835}, 
Dec=$-$03{\deg}05{\am}\dms{51}{\prime\prime}{34}. 
 An enhancement in the X-ray emission is located just to the south of PS1 only separated from the best-fit position of this source by approximately \dms{13}{\prime\prime}{5}. The sources PS1 and PS2 are respectively catalogued as CXO~J18443.4$-$030520 and CXO~J18444.1$-$030549 in the {\Chandra} Source catalogue \citep[CSC,][]{evans10}, and as 3XMM~J184442.9$-$030525 and 3XMM~J184441.9$-$030550 in the {\XMM} Serendipitous Source catalogue \citep[3XMM-DR4,][]{watson09}. 
We used the NASA/IPAC Extragalactic Database (NED) to search for feasible associations of the two sources detected in our {\Chandra} image with extragalactic objects. We found two NED sources separated from PS1 and PS2 by \dms{2}{\prime\prime}{8} and 3{\as}, respectively. These angular distances are larger than the average pointing uncertainty of {\Chandra} ($\leqslant$ 1{\as}) and thus the X-ray sources cannot be directly associated with the NED objects.
We then analyzed possible near-infrared identifications in the Two Micron All Sky Survey (2MASS) All-Sky catalogue of Point Sources \citep{cut03}, which has an astrometry uncertainty better than \dms{0}{\prime\prime}{1}. The search was done within a region of 2{\as} to account for errors in the determination of the X-ray positions and to minimize the probability of chance coincidence. We found a possible infrared (IR) counterpart located about $\sim$1{\as} from 
the X-ray source PS2, but no IR point sources were distinguished within a matching radius of $\sim$4{\as} around the X-ray source PS1.  Examining the Digital Sky Survey (DSS), no obvious optical correlations were identified in association with the X-ray point sources within a 5{\as} radius.

\subsubsection{X-ray spectroscopy}
\label{secc_spec_X}
We used the CIAO tool {\ttfamily specextract} to obtain the {\it Chandra} spectra, and the SAS {\ttfamily evselect} task for {\XMM} spectra. For all observations, the backgrounds were selected from circular regions free of diffuse emission and point sources. For each region, we extracted individual spectra from the data reported in Table~\ref{table_X_1}, thus obtaining two spectra for {\it Chandra} (corresponding to observations \#11232 and \#11801) and six spectra for {\XMM} (corresponding to the MOS1, MOS2, and PN cameras of observations \#0602350101 and \#0602350201).
Then, we combined the spectra to produce a single merged spectrum for {\it Chandra} and a single merged spectrum for {\XMM}, using the CIAO tool {\ttfamily combine$_{-}$spectra} and the SAS task {\ttfamily epicspeccombine}, respectively. During the fitting process, we checked the consistency of the results obtained from the simultaneous fit of the two merged spectra and the results of the simultaneous fit of all the eight spectra. We find that the model parameters always agree and are better constrained when fitting the merged spectra (i.e. confidence ranges are smaller).

\begin{figure*}[ht!]
\centering
\includegraphics[scale=0.8]{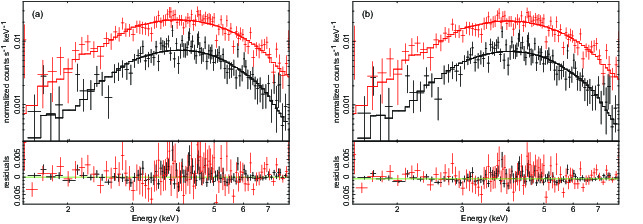}
  \caption{X-ray spectrum of G29.37+0.1 in the 1.5-8.0~keV energy band. We show the merged {\it Chandra} (in black) and {\XMM} (in red) spectra obtained combining the individual spectra for the different observations, as described in the text. The lines are the best-fit with \bf a) \rm an absorbed power-law model and \bf b) \rm
an absorbed bremsstrahlung model. } 
  \label{fig_X3_extragalactic}
\end{figure*}

After removing the contribution from the point sources PS1 and PS2, the spectrum of the nebula 
was extracted from the elliptical region of 65{\as} $\times$ 41{\as} centred at 
RA$\sim$18$^{\mathrm{h}}$44$^{\mathrm{m}}$43$^{\mathrm{s}}$,
Dec$\sim$-03$^{\circ}$05$^{\prime}$\dms{39}{\prime\prime}{1} shown in the inset of Fig.~\ref{fig_X1}.
The individual merged {\Chandra} and {\XMM} spectra were 
grouped with a minimum of 20 and 40 counts per bin, respectively. 
The two merged spectra were fitted simultaneously in the 1.5-8.0~keV band using Xspec 
(version 12.8.1) and $\chi^2$ statistics. 
The model parameters of the \it Chandra \rm and \it XMM-Newton \rm dataset were tied together and only the normalizations were allowed to
vary independently to each other to account for different instrumental calibrations between the two telescopes.
The merged {\Chandra} and {\XMM} spectra are shown in Fig.~\ref{fig_X3_extragalactic}. There were 
2050 and 3573 counts in the 1.5-8.0~keV band for the {\Chandra} and {\XMM} spectrum, respectively.
The X-ray spectrum of {\G29}  appears as a continuum with no hints of the presence of emission lines.

We fitted the spectrum of the diffuse X-ray emission considering different scenarios.
If the X rays are produced in the presumed radio galaxy, the emission may be originated by electrons accelerated at the lobe
that radiate  through a non-thermal mechanism (synchrotron or inverse Compton) or it may be thermal emission produced
by shocked gas surrounding the lobe. Then, we used both a power-law and a bremsstrahlung model to fit the non-thermal and thermal emission, 
respectively. In our procedure, we included a photoelectric absorption component to model the absorption of the Galactic
interstellar medium. The best-fitting model parameters are summarized in Table~\ref{table_X_diffuse}. The listed errors correspond
to the 90\% confidence level.

In the case of an absorbed power-law model ($wabs \times power law$ 
in the Xspec environment), letting the hydrogen column density $N_{\mathrm{H}}$ and the 
photon index $\Gamma_X$  vary freely, the fit resulted in a good 
$\chi^{2}_{dof}$=$\chi^{2}$/$dof$$\sim$~1.05 
($dof$ being the number of degrees of freedom), a Galactic absorption column density of
$N_{\mathrm{H}}(\mathrm{Gal})= 9.58 \times 10^{22}$~\cm{-2}, and a photon index $\Gamma_{X} = 1.76$. 
After correcting for Galactic absorption, the flux\footnote{Hereafter, to derive physical properties of the X-ray emission, we use the flux obtained from the
\it Chandra \rm dataset. Using the flux of the \it XMM-Newton \rm 
observations yields similar results.} in the 1.5-8.0~keV band is 
$\sim$1.6$\times10^{-12}$~erg~cm$^{-2}$~s$^{-1}$. 
For an assumed redshift $z=$0.05 of the radio emission lobe, we obtain an absorption-corrected luminosity in the 1.5-8.0~keV band of 
$\sim$9.1$\times$10$^{42}$~erg~s$^{-1}$, while for $z$=0.01 and $z$=1 the estimated luminosity are
$\sim$3.4$\times$10$^{41}$~erg~s$^{-1}$ and $\sim$8.2$\times$10$^{45}$~erg~s$^{-1}$, respectively 
(we present in Sec.~\ref{redshift} a further detailed 
discussion on the redshift estimate to the potential radio galaxy G29.37+0.1).

On the other side, in the case of modelling the X-ray emission with a pure thermal bremsstrahlung component 
($wabs\times zbremms$ in the Xspec environment), for $z$=0.05 the resultant 
best-fit hydrogen column density and electron temperature ($T_{\mathrm{e}}$) are 
$\sim$8.9$\times$10$^{22}$~cm$^{-2}$ and
$T_{\mathrm{e}}$$\sim$14~keV, respectively. The flux in the 1.5-8.0~keV band (corrected for absorption) 
is $\sim$1.4$\times$10$^{-12}$~erg~cm$^{-2}$~s$^{-1}$. 
In this model, the temperature is not well constrained ($T_{\mathrm{e}}$$\sim$8-34~keV for a 
90\% confidence level) and the 
best-fit value ($T_{\mathrm{e}}$$\sim$14~keV) is much
higher than the temperatures measured toward the lobes of other 
radio galaxies for which a thermal model works well (\citealt{grandi03,isobe05}). 
Hence, we favour a non-thermal origin for the X-ray emission.

We cannot omit that the obtained absorption parameter for both the non-thermal and thermal models 
($N_{\mathrm{H}}(\mathrm{Gal})$$\sim$9$\times$10$^{22}$~cm$^{-2}$)
is high compared with the Galactic value $\sim$2$\times$10$^{22}$~cm$^{-2}$  measured  from HI emission 
toward RA$\sim$18$^{\mathrm{h}}$44$^{\mathrm{m}}$\dms{42}{\mathrm{s}}{3},
Dec$\sim$$-$03$^{\circ}$06$^{\prime}$23$^{\prime\prime}$ (or $l$$\sim$\dms{29}{\circ}{4}
$b$$\sim$\dms{0}{\circ}{}08), the central coordinates of the northeastern lobe  
\citep{kalberla05}. 
We also tried to fit the X-ray spectrum keeping fixed the hydrogen
column density in 2$\times$10$^{22}$~cm$^{-2}$, but we were unable to obtain statistically 
acceptable fits ($\chi^{2}_{dof}>2$) for both non-thermal and thermal models. 
In previous studies of the diffuse X-ray emission from the lobes of radio galaxies, the values 
of the hydrogen column densities obtained from the fitting are generally similar to the 
Galactic ones derived from the emission of the HI at 21~cm, typically 
$N_{\mathrm{H}}(\mathrm{Gal})\lesssim 10^{21}$~cm$^{-2}$ 
(see for instance \citealt{kraft03,isobe06,migliori07}). 
However, it is noticeable that in these cases the sources are located well outside the Galactic plane, while the line of sight in the direction of G29.37+0.1 passes through the Galactic disk. 

\begin{table*}[ht]
\caption{Best-fit parameters for the diffuse X-ray emission in G29.37+0.1 between 1.5 and 8.0~keV. 
The reduced $\chi^{2}$ ($=\chi^{2}/dof$, where $dof$ are the degrees of freedom) is $\chi^{2}_{dof}$, while
$F$ is the absorption-corrected flux in the 1.5-8.0~keV energy band. 
We report the fluxes of \it Chandra \rm $F_{Chandra}$ and \it XMM-Newton \rm $F_{XMM}$ dataset separately.}
\centering
\setlength{\extrarowheight}{5pt}
\begin{tabular}{ccccccc}
\hline\hline
Model     & $\chi_{dof}^{2}$($dof$) & $N_{H}$            & $\Gamma_{X}$      & $T_{e}$  & $F_{Chandra}$     & $F_{XMM}$ \\
          &                     & [$\times10^{22}$~cm$^{-2}$] &               & [keV]    &   
\multicolumn{2}{c}{[10$^{-12}$~erg~cm$^{-2}$~s$^{-1}$]}                      \\
\hline
Power law & 1.05(172)        & 9.58$_{-1.32}^{+1.51}$  & 1.76$_{-0.29}^{+0.31}$ & -   &  $1.56\pm0.08$  & $1.20\pm0.05$  \\
Bremsstrahlung  & 1.05(172)  & 8.96$_{-1.03}^{+1.16}$  & - & 13.73$_{-6.05}^{+19.51}$ &  $1.42\pm0.08$  & $1.10\pm0.05$  \\
\hline
\label{table_X_diffuse}
\small
\end{tabular}
\end{table*}

The apparent correlation between X-ray and radio emission may, however, simply be due to a superposition along the line of sight rather than
an X-ray counterpart to the radio emitting plasma in our candidate galaxy. Therefore, as an alternative analysis to the 
extragalactic scenario, we point out the possibility that the X-ray emission arises from our Galaxy.
We found that fitting the spectrum with a power-law model yields a photon index 
$\Gamma_{\mathrm{x}}$$\sim$1.76 (Table~\ref{table_X_diffuse}) compatible with that estimated in pulsar wind nebulae \citep{kargal10}, this being another possible origin for the X-ray emission.

\begin{figure}[h]
  \centering
\includegraphics[width=7cm]{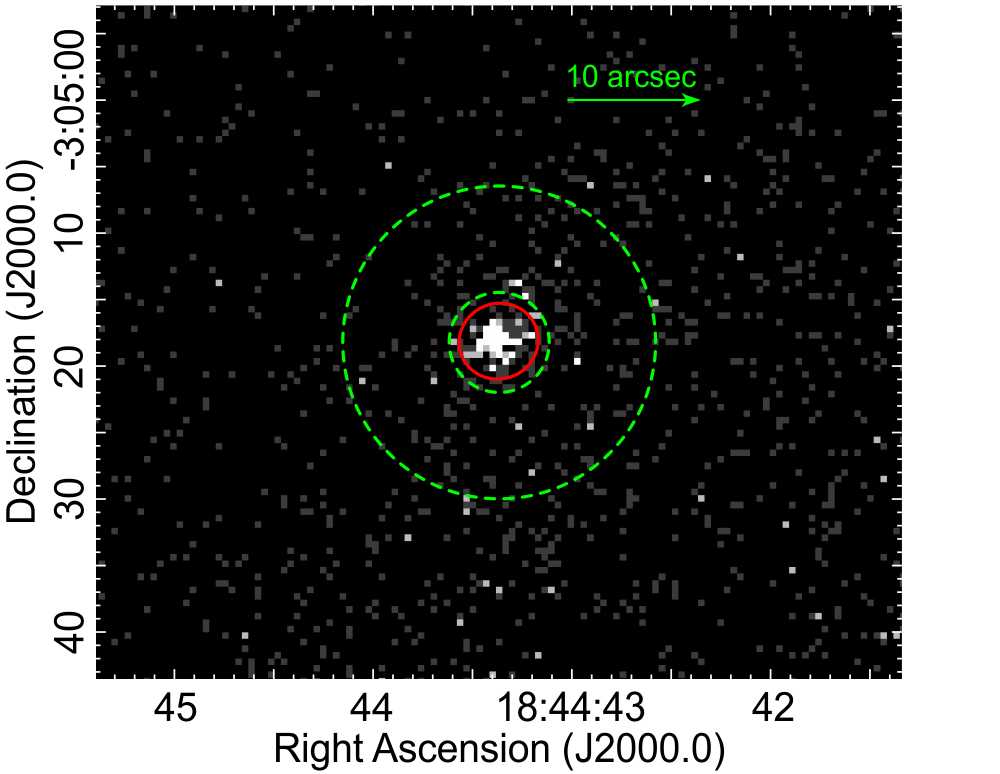}
  \caption{ACIS-I full-resolution image toward the point source PS1 in the 1.5$-$7.5~keV energy band, obtained from the combination of {\Chandra} observations \#11232 and \#11801. The red circle and the green annulus are the source and background regions, respectively. 
}
  \label{PS1_img}
\end{figure}

\begin{figure}[h]
\centering
\includegraphics[width=7cm]{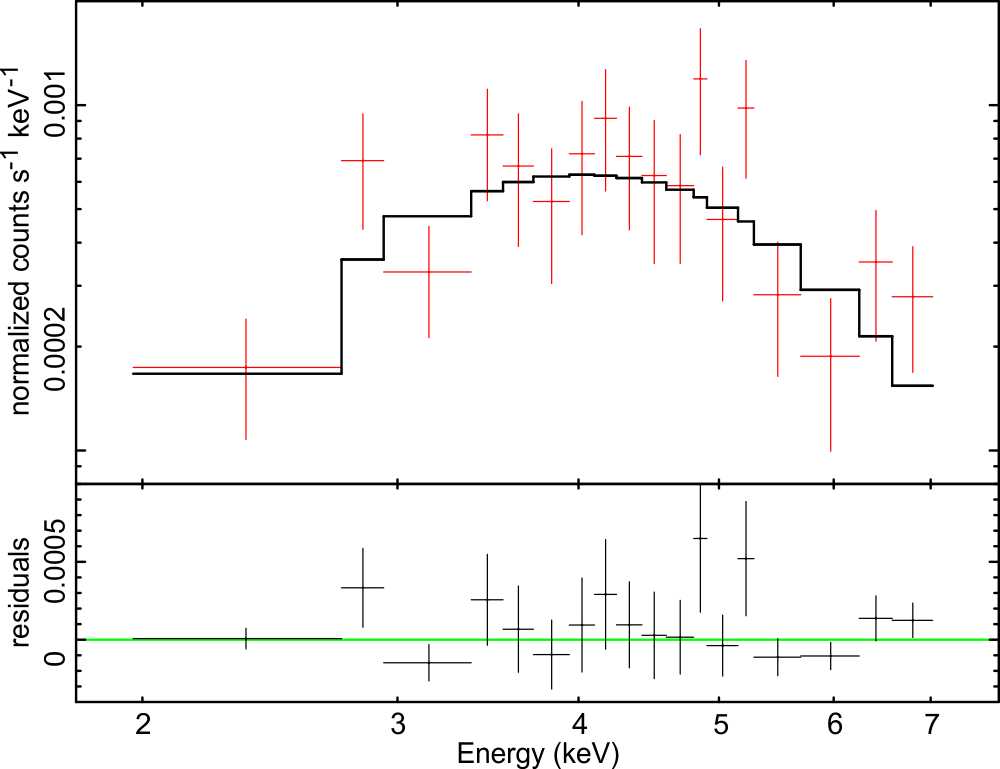}
  \caption{ACIS-I spectrum of the point source PS1, obtained from the combination of observations \#11232 and \#11801. The solid line is the best fit power law model with a column hydrogen of $9.58\times10^{22}$~\cm{-2}.}
  \label{PS1_spec}
\end{figure}

As mentioned above, the X-ray compact sources PS1 and PS2 are superimposed on the 
diffuse emission.
In the next paragraphs, we perform a spectral analysis of their X-ray emission to determine if, in a PWN scenario, they are reliable 
candidates to be the pulsar which powers the high-energy nebula. We only use \it Chandra \rm observations to take advantage of the small 
extent of its PSF, suitable for spectral analysis of point sources.
In Fig.~\ref{PS1_img}, we show a full-resolution image of the X-ray emission toward PS1, obtained from the combination of the \it Chandra \rm datasets. 
Individual spectra of both {\Chandra} observations were combined into a single merged spectrum, which was grouped with a minimum of eight~counts per bin (see Fig.~\ref{PS1_spec}). We obtained 148~counts in the 1.5-7.5~keV band. In spite of the low number of counts, the spectral analysis can still provide some insights about the nature of PS1.

The X-ray emission from pulsars may correspond either to thermal emission from the neutron star surface or to non-thermal emission from the magnetosphere. The thermal emission is usually well described by a black body model with a characteristic temperature of $\sim$0.1~keV, while the non-thermal spectrum is well described by a power-law model with photon index in the range $1.0 \lesssim \Gamma_{X} \lesssim 2.0$ \citep{kargal10}. 
We fitted the spectrum of PS1 by considering the absorption column density fixed to the value obtained for the diffuse emission ($9.58\times10^{22}$~\cm{-2}) and as a free parameter. 
In Table~\ref{PS1_table} we report the best-fit values obtained using $\chi^{2}$ statistics. Due to the low number of counts per bin, we also
fitted the spectrum using Cash statistics and found similar parameter values.

\begin{table*}[h]
\caption{Best-fit parameters of the spectral analysis of point source PS1. The symbol (*) indicates that the parameter was frozen.}
\label{PS1_table}
\small
\centering
\setlength{\extrarowheight}{5pt}
\begin{tabular}{ccccc}\hline\hline
 \multirow{2}{*}{Model}  &  \multirow{2}{*}{{$\chi$$^{2}_{dof}$}($dof$)}  &           {\NH}             &  \multirow{2}{*}{$\Gamma_X$} &          
$T_e$           \\      
   &                                      &  [$\times10^{22}$~\cm{-2}]  &                              &          [keV]    \\ \hline 
 Power law          &         0.86(16)             &           9.58*             &    $1.75_{-0.65}^{+0.63}$    &   --  \\
 Power law         &                0.91(15)             &  $9.31_{-6.73}^{+10.83}$    &    $1.69_{-1.66}^{+2.15}$   & --              \\
 Black body        &                0.95(16)             &           9.58*             &   --   &  $1.17_{-0.21}^{+0.35}$ \\                  
 Black body        &                1.01(15)             &  $5.85_{-4.62}^{ +7.42}$    &               --             &  $1.51_{-0.57}^{+1.86}$  \\\hline                     
\end{tabular}
\end{table*}

From the results reported in Table~\ref{PS1_table}, we conclude that both power-law and  black body models provide similar adequate fits. For the power-law, when {\NH} is frozen to 
9.58$\times$10$^{22}$~\cm{-2}, the resulting photon index $\Gamma_X$$\sim$1.75 is compatible with the values measured toward other pulsars powering PWNe \citep{kargal10}. The absorption-corrected flux in the 1.5-8.0~keV band
is 1.34$\times$10$^{-13}$~erg~s$^{-1}$~cm$^{-2}$. When {\NH} is a free parameter, its best-fit value is similar to the value toward the diffuse emission, but the errors are larger. In this case, $\Gamma_X$ is also badly constrained.
Regarding the thermal model (black body), the obtained electron temperatures $T_e$ for both {\NH}, frozen and free, are $\sim$1~keV, a value higher than that expected in a young pulsar (PSR). Fixing $T_{\mathrm{e}}=0.1$~keV results in an unacceptable fit ($\chi^{2}_{dof}$$\gg$1) for both {\NH} fixed to $9.58\times10^{22}$~\cm{-2} and free. Thus, if PS1 is the compact object powering the diffuse emission detected in the direction of {\G29}, its nature is likely non-thermal. 

In relation to PS2, this point source presents a variable count rate between the two 
\it Chandra \rm observations, increasing from 
$\sim$0.6$\times$10$^{-2}$~cts/s (Obs-ID 11232) to $\sim$2.1$\times$10$^{-2}$~cts/s (Obs-ID 11801) 
in the 0.7-8.0~keV energy band. For this reason, we fitted both observations separately. We extracted spectra from circular regions of 2\as centred in PS2 and we fitted them with non-thermal (power-law) and 
thermal (blackbody) models, as we did for PS1. For observation 11232, we obtained 190 counts in the 
0.7-8.0~keV band and the power-law model yields a column density 
$N_{\mathrm{H}}$$\sim$0.2$\times$10$^{22}$~cm$^{-2}$ and a spectral index 
$\Gamma_X$$\sim$$0.78$. 
For observation 11801, we obtained 611 counts, $N_{\mathrm{H}}$$\sim$0.4$\times$10$^{22}$~cm$^{-2}$ and 
$\Gamma_X$$\sim$$0.95$. 
The black body fit results in extremely low hydrogen column densities ($N_{\mathrm{H}}$$\sim$10$^{13}$~cm$^{-2}$) 
and electron temperatures of $\sim$1.3~keV for both observations. 
We obtained unacceptable fits when we fixed the hydrogen column density to the best-fit value derived for the diffuse emission ($N_{\mathrm{H}}$=9.58$\times$10$^{22}$~cm$^{-2}$) for both models.
Thus, based on the large discrepancy between the column densities derived for the diffuse emission 
and for PS2, we argue that this point source is 
not related to the X-ray nebula and conclude that in a PWN scenario PS1 
could be the putative pulsar powering the X-ray diffuse emission toward G29.37+0.1. 

\section{Discussion}
\label{discussion}
In this section, we put together all the pieces of information that we have collected in order to 
constrain the physical nature of the continuum emission from the puzzling radio source G29.37+0.1.  
Apart from the radio and X-ray analysis presented in the previous sections,
here we examine two basic hypotheses. Firstly, we consider the possibility 
that the observed radio emission  is from an extragalactic object. 
Secondly, we discuss, on the basis of
the distribution of both molecular and atomic gas located in projection adjacent 
to G29.37+0.1, the alternative idea in which the observed radio and X-ray emission
do not originate in a single object but from superimposed structures. 

\subsection{Extragalactic emission from G29.37+0.1?}
\label{redshift}

\subsubsection{The S-like feature in G29.37+0.1}
\begin{figure}[!ht]
 \centering
\includegraphics[width=8cm]{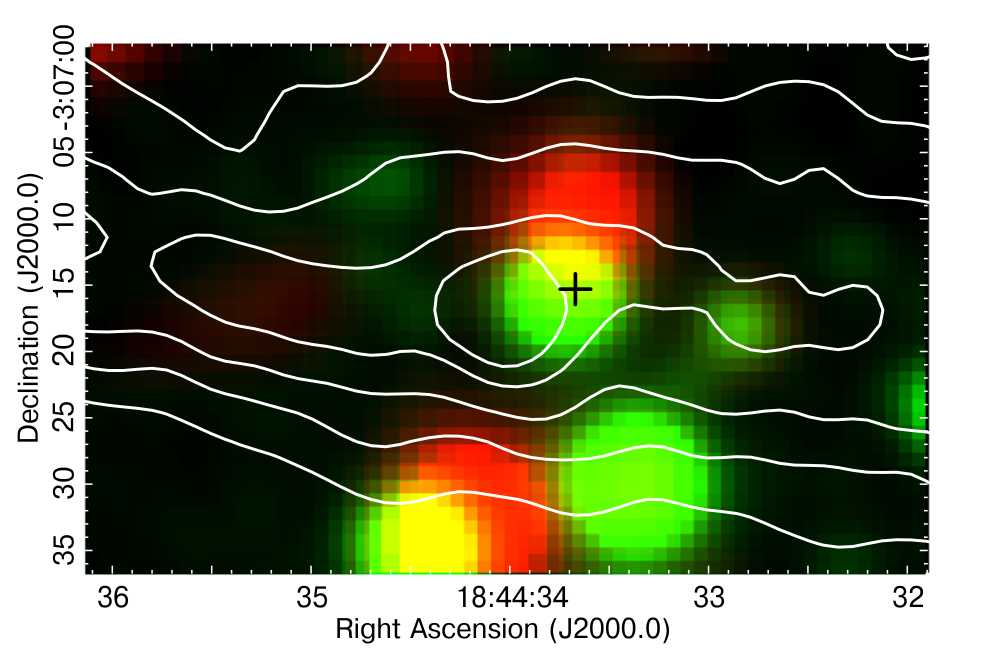}
 \caption{Close-up vision of the central region in G29.37+0.1. 
The radio emission at 610~MHz is shown as contours, starting at nine times that of the 
local noise and increasing by a factor of 3.3. The radio contours are overlaid on the 3.4~$\mu$m infrared (W1 band, in red) and optical (in green) fields 
taken from WISE (FWHM$\sim$6$^{\prime\prime}$) and DSS  (resolution 1$^{\prime\prime}$/pixel). 
The position of the 2MASS object J18443367$-$0307153, the possible infrared counterpart to both the DSS optical and WISE emission, 
is marked with a plus symbol. 
}
\label{center}
\end{figure}

As mentioned above, the observed surface brightness distribution of the S-like feature in
G29.37+0.1 is found to be similar to a radio galaxy.
Morphological properties, however, are often not sufficient by themselves to classify an
extended source as a galaxy, especially if it has a faint magnitude.
Classification methods based on infrared colour-colour plots  along with the search for optical counterparts are widely used for
this purpose \citep{macha01,pollo10,kovacs15}. 
Using the DSS we have identified the radio core
with an optical source at the position 
RA$\sim$18$^{\mathrm{h}}$44$^{\mathrm{m}}$\dms{33}{\mathrm{s}}{6}, 
Dec$\sim$$-$03$^{\circ}$07$^{\prime}$\dms{15}{\prime\prime}{5}. Its optical
magnitudes are b$\simeq$15.1, r$\simeq$13.8, i$\simeq$12.3, and 
v$\simeq$13.6 \citep{mon03,zac04}. This optical counterpart is very probably determined as 
a non-stellar object in the USNO-B1.0 catalogue \citep{mon03}. 
We also searched for an infrared counterpart to the optical source using 
near- and mid-infrared photometry and imaging extracted from 2MASS \citep{skr06}. We identified 
a 2MASS point source, named J18443367$-$0307153, lying in the region of the DSS source. 
No redshift information is available for the identified optical counterpart. 
Thus, if we assume that 
the DSS and 2MASS are counterparts to the radio core of G29.37+0.1, 
and take the well-known infrared K-band versus \it z \rm 
relation for radio galaxies into account \citep{willott03}, the measured K$\simeq$10.86~mag
for the 2MASS source implies a redshift $z$$\sim$0.05 for our radio galaxy candidate.
In order to gain additional insight into the nature of the nucleus component in
G29.37+0.1, we used infrared data extracted from the 
All-Sky Wide-field Infrared Survey Explorer (AllWISE) Source Catalogue
\citep{mainzer11}. The four WISE bands from W1 to W4 are centred at wavelengths of 3.4, 4.6, 12, and 22~$\mu$m, respectively. 
In the WISE field, the source J184433.65$-$030710.0 is the closest to the nuclear radio component of G29.37+0.1 and hence to its proposed optical counterpart (with an half power beam widht (HPBW) of $\sim$6$^{\prime\prime}$ similar to
the WISE W1-data, the accuracy of the GMRT image is about 2$^{\prime\prime}$, i.e. 1/3 of its angular resolution). 
The global magnitudes of the WISE source 
are W1=8.852$\pm$0.023,
W2=8.793$\pm$0.023, W3=9.168$\pm$0.397, and W4=7.071$\pm$0.522.
Figure~\ref{center} depicts the 610~MHz radio emission from the central region of
G29.37+0.1, illustrated with brightness contours overlaid onto a WISE 3.4~$\mu$m 
(W1-band) and DSS optical images. 
Despite the projected angular offset between the central position of the WISE source 
and the DSS/2MASS object, if we still consider the WISE source as a plausible counterpart, 
we could determine the nature of the presumed host galaxy 
by combining the photometric information of WISE with 2MASS data. 
Following \citet{kovacs15}, we employed the 
most sensitive WISE band W1 and the \it J \rm band from 2MASS to efficiently separate 
elliptical galaxies from stars. 
By applying their W1$-$J$\leq$$-$1.7 colour cut, which guarantees the lowest stellar contamination, we found 
that the 2MASS source J18443367$-$0307153 lies in the galaxy
locus of the W1 vs. W1$-$J colour-magnitude plane. 
In this case, by linking the nuclear component observed at radio wavelengths to
the brightest S-shaped morphology in G29.37+0.1, an extragalactic origin for the 
radio emission observed from this whole structure seems plausible. At this point, it is also necessary 
to mention that a second 2MASS object (J18443361$-$0307095) was also found lying at the centre of the WISE
source. In our analysis, we did not consider it to be related to the 
nuclear emission in G29.37+0.1 due to the lack of an additional optical identification. 

In what follows, we estimate a set of 
basic physical parameters for G29.37+0.1, including the radio luminosity, projected linear size, 
equipartition magnetic field, and energy density, by adopting the obtained $z$ value. 
However, we recognize that due to the low K-magnitude of the identified 2MASS source, which is somewhat
marginal in the K-\it z \rm relation of \citet{willott03},
along with our uncertainty in classifying it as the counterpart to the radio core, it is not 
possible to provide a reliable constraint for the redshift. Hence, the same physical magnitudes were also estimated by adopting an interval 
0.01$<z<$1 of probable values for the redshift at G29.37+0.1. This range is consistent with those derived from previous studies of radio galaxies 
and giant radio galaxies (with projected sizes over Mpc), for example \citet{lar01,willott03}. Using $z=$0.05, the projected linear size of the 6-arcmin S-shaped structure is
373~kpc \citep[if H$_{0}$=71~km~s$^{-1}$~Mpc$^{-1}$, $\Omega_{\mathrm{m}}$=0.27, and
$\Omega_{\mathrm{\Lambda}}$=0.73 are adopted,][]{spe03}, or between 74~kpc and 8~Mpc
if a redshift range of 0.01$<z<$1 is considered. In the latter case, G29.37+0.1 would belong to the 
class of giant radio galaxies. 
At $z$=0.05 the total 610-MHz luminosity [W~Hz$^{-1}$] has a log value of 25, or 
it ranges between 23.5$<\mathrm{log}\,L_{610}<$27.5 for 0.01$<z<$1. Using our estimate for the radio spectral index ($\alpha_{\mathrm{R}}$$\sim$0.6), the corresponding logarithmic radio power at 1.4~GHz is 24.8 at $z$=0.05 
(or 23.3$<\mathrm{log}\,L_{\mathrm{1400}}<$27.4, 0.01$<z<$1).
Our results are in broad agreement with those presented by \citet{macha06}, where a
range  of total power 24.3$<\mathrm{log}\,L_{\mathrm{1400}}[\mathrm{W~Hz^{-1}}]<$28.6 is measured for 
a redshift interval from 0.03 to 1.8 in a large sample of selected normal FRII radio sources. 
We are aware that for redshift values greater 
than $\sim$0.14 (for which the linear size of the S-shaped structure is larger than 1~Mpc), the radio 
luminosity at 1400~MHz ($\mathrm{log}\,L_{\mathrm{1400}}$$\sim$25.6) falls in the region of 
transition between FRI and FRII-type large radio 
galaxies \citep[see][]{lar04}.

We also estimate the magnetic field strength, B$_ {\mathrm{eq}}$, averaged over the whole S-shaped 
volume and over each lobe separately. To do this calculation, we consider the minimum 
energy condition, which roughly implies equipartition of energy densities between
fields and relativistic particles. Following the revised formalism proposed by
\citet{beck05}, we assume a filling factor of unity corresponding to a radio source
completely filled by the magnetic field. 
Regarding the $K_{\mathrm{0}}$ parameter in their equation (A18), which represents the ratio of the 
number density of protons to that of electrons per particle energy interval, 
we consider $K_{\mathrm{0}}=(\frac{m_{\mathrm{p}}}{m_{\mathrm{e}}})^{\alpha}$$\sim$100 
(see also Eq.~7 in \citealt{beck05}'s work). This value, computed by adopting 
$\alpha$$\approx$$\alpha_{\mathrm{inj}}$$\sim$0.6 (where $\alpha_{\mathrm{inj}}$ is the 
injection index of the accelerated particles) is in concordance with particle energy content dominated by 
the protons.

\begin{table*}[h!]
\centering
\caption{Equipartition parameters for the bright components in the radio source G29.37+0.1.}
\begin{tabular}{cccccc}                                               \hline\hline
 Component & z & B$_{\mathrm{eq}}$ & u$_{\mathrm{eq}}$ &$\mathrm{log}\,L_{\mathrm{1400}}$ \\
           &   & [nT]              & [10$^{-14}$~J~m$^{-3}$] \\\hline
   S-shape &  0.05   & 0.76            & 53.8 & 24.75        \\
     &  0.01-1 & 1.20-0.32 & 132.6-9.4     &    23.35-27.37      \\ \hline
   Northeastern  &  0.05 & 0.75 & 52.0 & 24.46  \\ 
     lobe               & 0.01-1 &  1.17-0.31 & 127.8-9.1 & 23.06-27.08 \\ \hline
   Southwestern  & 0.05 & 0.72 & 47.53 & 24.1 \\
       lobe             & 0.01-1 & 1.12-0.30 & 116.8-8.3 & 22.7-26.7 \\ \hline
\label{equipartition}
\end{tabular}
\end{table*}

In our calculation, we have assumed a cylindrical geometry for the lobes. The size of the 
lobes was estimated from the lowest radio contour drawn in Fig.~\ref{contours} with the base 
diameters equal to the average width of each lobe,
\dms{1}{\prime}{3} and 1$^{\prime}$, and lengths 
of \dms{2}{\prime}{6}  and \dms{2}{\prime}{2} for the 
northeastern and southwestern structures, respectively.
The total volume of the whole
S-shaped structure was determined from the sum of the individual lobe's volume. The global
equipartition parameters estimated for the whole presumed radio galaxy 
(i.e. the S-shaped feature) and for each lobe in G29.37+0.1 are summarized in 
Table~\ref{equipartition}. 
It should be pointed out that the spectral break suggested in the radio continuum spectra of both 
lobes (Fig.~\ref{spectrum}) prevents the use of a global spectral index for each lobe. Therefore, 
in determining the physical parameters of the two radio lobes in G29.37+0.1 we used  
an intermediate spectral index $\sim$0.5, a compromise between the values measured 
between 610 and 1400~MHz in both features. 
Firstly, we found that similar values of the equipartition magnetic field and energy density were reported in the literature for 
other FRII radio galaxies also observed with GMRT \citep{har16}, as well as for a  sample of giant radio galaxies \citep{konar08}.
Considering a smaller value for $K_{\mathrm{0}}$ yields magnetic fields up to
three times smaller than those computed assuming dominance of protons ($K_{\mathrm{0}}$$\gg$1). 
We, however, emphasize that these results should be read with caution due to the 
uncertainties inherent in the redshift determination and in fundamental parameters related to
the energy content of the radio emission (e.g. K$_{0}$, the volume filling factor, or the extent of the source,
see \citealt{har16} for a further discussion regarding departure from the equipartition approach). 
Furthermore, it should not be ignored that there is scarce evidence that equipartition with a magnetic 
field uniformly distributed in the emitting region 
actually takes place in lobes of radio galaxies.

\subsubsection{X-ray associated with the radio-lobes in G29.37+0.1?}
We showed in Sec.~\ref{secc_spec_X}  that the 
most plausible origin of the diffuse X-ray emission is non-thermal. 
To investigate if the synchrotron process responsible for the radio emission could also produce 
the observed X-rays, we estimated the X-ray flux predicted by the extrapolation of the radio 
emission to the keV domain.
We used the relation $F_{\nu}\propto \nu^{-\alpha}$ and took the spectral radio index 
$\alpha_{\mathrm{R}}=0.47$, as derived in 
Sec.\ref{radio-spectrum} 
for the northeastern radio lobe. For a reference energy of 1 keV ($\nu\sim2.4\times10^{17}$ Hz), 
we obtained $F_{1\mathrm{keV}}$$\sim$10$^{-5}$~Jy, 
much higher than the X-ray flux reported in Table~\ref{table_X_diffuse} for the power-law fit 
($\sim$9.9$\times$10$^{-8}$~Jy in the 
1.5-8.0~keV energy band).
Thus a simple power-law distribution for the electron population cannot account simultaneously 
for the radio and X-ray emissions. 
However, we note that the X-ray spectral index $\alpha_X$=$\Gamma_X-1$$\sim$0.76 is harder than 
$\alpha_{\mathrm{R}}=0.47$, so both emissions 
can still be connected if we invoke a spectral break between the radio and the keV bands.

Now we explore the inverse Compton (IC) mechanism as responsible for the emission in the X-ray domain. 
There are several sources of photons that may be IC-scattered up to X-ray energies in radio galaxies. 
In high-density electron environments (such as hotspots) the dominant photon population
comes from the synchrotron emission and leads to the so-called synchrotron self-Compton mechanism (SSC, \citealt{hardcastle04}). 
In the low-density lobes of radio galaxies other sources of seed photons are required. 
The photons from AGN of the galaxy and the cosmic microwave background (CMB) can provide a field to be scattered to keV energies 
\citep{croston05}.
A comparison between the energy density of the field of photons from the synchrotron emission, the central AGN, and the CMB
has been used to discriminate between different IC scenarios in the lobes of radio galaxies (see \citealt{isobe02,grandi03,isobe05}). 
Such studies have shown that in FRII sources the CMB is likely to be the main source of low energy photons for the IC process.

If X-ray emission in the northeastern lobe is caused by IC scattering of CMB photons, we can estimate the magnetic field $B$. 
We follow \citet{harris79}, who derived an expression for the magnetic field in terms of the the radio continuum spectral index $\alpha$ and 
the radio synchrotron, $F_{\mathrm{r}}$, and X-ray, $F_{\mathrm{X}}$, fluxes (adopting 
the notation of the authors): 

\begin{equation}
B=A(\alpha)\left(1+z\right)^{\frac{\alpha+3}{\alpha+1}}\left(\frac{F_r}{F_X}\right)^{\frac{1}{\alpha+1}}\left(\frac{\nu_\mathrm{r}}{\nu_\mathrm{X}}\right)^{\frac{\alpha}{\alpha+1}},
\label{BIC}
\end{equation}

\noindent 
where $A(\alpha)=\left[10^{-47}\left(5.05\times10^4\right)^{\alpha}C(\alpha)G(\alpha)\right]^{\frac{1}{\alpha+1}}$. 
Here, $C(\alpha)$ can be approximated with a value of 1.15$\times$10$^{31}$ for 0.5$<\alpha<$2.0. The factor $G(\alpha)$ 
is a correction to account
for the IC spectrum produced by the scattering of electrons on the entire black body distribution of CMB photons with
a temperature $2.7(1+z)$~K, 
instead of considering a monochromatic CMB with 2.7~K photons. The value of $G(\alpha)$ was numerically estimated
by \citet{harris79} for different values of $\alpha$. 
Taking the spectral index estimated from radio observations for the lobes in G29.37+0.1 
($\alpha=\alpha_{\mathrm{R}}=0.47$), 
the corresponding  correction factor is $G(\alpha)$=$\sim$0.5. 
The magnetic field $B$ 
calculated from the above equation with the reported values of $C(\alpha)$ and $G(\alpha)$ is expressed in Gauss. 
We took the radio flux at $\nu_{\mathrm{R}}$=1400~MHz for the northeastern 
lobe F$_{\mathrm{1400MHz}}$=0.54~Jy (see Table~\ref{S-fluxes}).
In the X-ray band, we used the absorption-corrected flux in the 1.5-8.0~keV from the best-fit power-law model, namely
F$_{X}$=9.9$\times$10$^{-8}$~Jy. Thus, from Eq.~\ref{BIC} we obtain B$\sim$0.0015~nT (or 0.015~$\mu$G) 
for z=0.05. The respective magnetic field in the 0.01$<z<$1 interval varies between $\sim$0.002 and $\sim$0.01~nT. 
As is evident, there is a substantial 
difference between the magnetic field strength 
constrained to the X-ray and radio flux data points
and that computed assuming the equipartition condition in G29.37+0.1 to be valid. 
Several previously published works have noted such a discrepancy (corresponding to 
$B\lesssim$ (0.3-2.5)$B_{\mathrm{eq}}$, e.g. \citealt{wil01,grandi03,har16}) 
in the lobes of radio galaxies from which the IC X-rays are detected. 
It is noteworthy that none of the previously published comparisons show differences 
as significant as that we have found.
As argued by \citet{carilli02}, one possibility to explain this issue 
may be an X-ray emission mainly originated in a thermal process rather than by inverse Compton radiation. 
We tested a composite model ($power law + zbremms$) and found that it produces an acceptable fit. 
A highly absorbed 
spectrum is encountered, with a column density $\sim$12$\times$10$^{22}$~cm$^{-2}$. The non-thermal component 
(with a best-fit photon index  
$\sim$2) dominates the entire spectrum, especially the hard band (E>2~keV). The thermal component (with a best-fit T$_{\mathrm{e}}$$\sim$0.3~keV)
contributes marginally in the lower energy band. Thus, the addition of a thermal component does not produce significant variations in the non-thermal
flux of G29.37+0.1 as compared with a simple power-law model.
Alternative explanations, to account for the differences in the determination of the magnetic field, 
include anisotropies in the pitch angle distribution of the cosmic ray electrons or 
magnetic field concentration in small filamentary structures, which leads to a reduced volume filling factor of the magnetic field (up to 10$^{-2}$) \citep[][and references therein]{beck05}.

Finally, we recall that the radio and X-ray emission may be produced by different sources superimposed along
the line of sight (namely, a radio galaxy and a galactic X-ray PWN). If this is the case, the 
estimation of the magnetic field using Eq.~\ref{BIC} is meaningless and this could account for the 
unexpected low values derived for this magnitude.

\subsection{Galactic emission from G29.37+0.1?}
While the morphological properties of G29.37+0.1 (mainly the S-like bright structure) themselves
seem well matched to the extragalactic hypothesis, due to 
its low Galactic latitude, a fortuitous projected positional superposition of other shell-type non-thermal Galactic sources, such as a supernova remnant, 
cannot be discarded.  
In this section, we attempt to put constraints on the kinematical distance 
to the continuum radio emission from G29.37+0.1 on the basis of hydrogen absorption. Additionally, we explore the molecular environment observed in projection toward the
radio source using $^{13}$CO data. 
Basically, we search for morphological and kinematic signatures
that could account for interstellar gas physically related to either the whole source or some of the
radio-synchrotron emitting components (especially its halo component).

\subsubsection{Distance constraints}
We analyze the kinematic distance to the observed radio synchrotron emission on 
the basis of neutral hydrogen emission and absorption profiles from the VGPS.
Useful spectra were obtained
in directions to the  northeastern and southwestern lobes and the nucleus, 
the three main components of the S-shaped feature. In the three cases, the emission 
spectrum was obtained toward 
regions where the radio continuum emission is bright, while the corresponding absorption profiles were obtained by subtracting the bright radio emission from nearby regions.
No HI spectra were obtained toward the halo due to the low surface brightness of this component. 

\begin{figure}[h]
\centering
  \includegraphics[width=\hsize,clip]{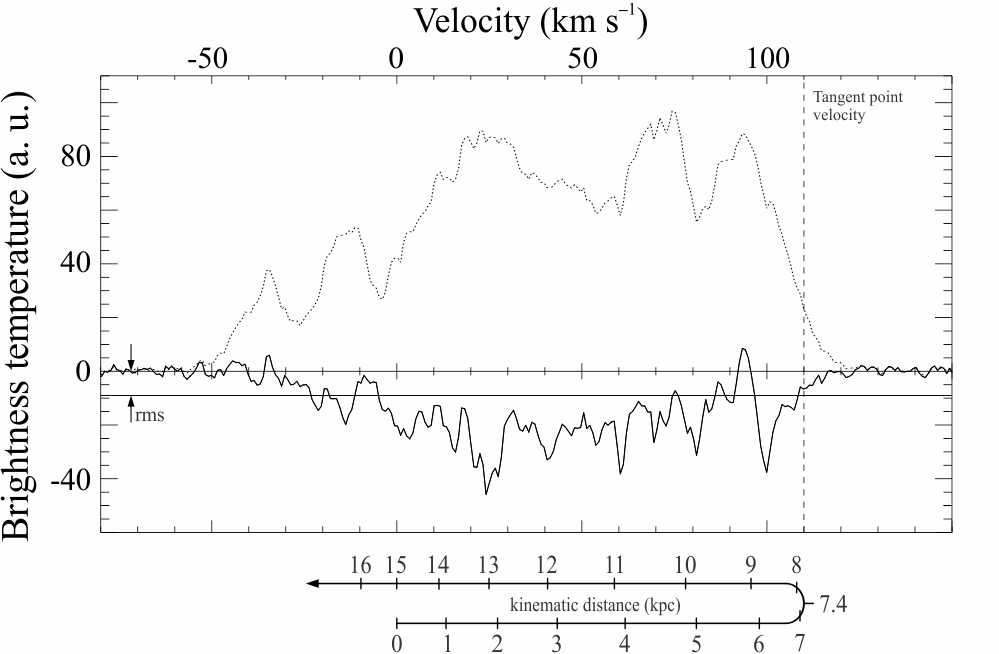}
\caption{HI emission (dotted line) and absorption (thick continuous line) spectra obtained toward the brightest (eastern) part of {\G29}. The thin horizontal lines indicate the rms noise for the absorption spectrum. 
Below, the assigned kinematic distances are denoted \citep{Fich-89}.
}
\label{SNR_HIabs_spectrum}
\end{figure}

In the direction of the northeastern lobe (see Fig.~\ref{SNR_HIabs_spectrum}) the HI absorption extends up to the
tangential point at 110~km~s$^{-1}$ (all the radial velocities mentioned in this paper
are measured with respect to the local standard of rest); the kinematic distance for this feature is 7.4~kpc 
\citep[converting velocities with the rotation model of the Milky Way of][with the Galactocentric distance 
R$_{0}$=8.5~kpc and the rotation velocity at the Sun $\Theta$=220~km~s$^{-1}$]{Fich-89}. 
In addition, at negative velocities HI absorption is observed roughly at $-$21 and $-$14~km~s$^{-1}$, 
although the signal-to-noise ratio is somewhat low. 
The kinematic distances corresponding to the negative features are 17.4 and 16.5~kpc, respectively. 
Taking all the HI spectra properties into account, we derive a lower distance limit 
d$\geq$17.4~kpc to the bright northeastern radio lobe of G29.37+0.1. 
Similar HI absorption and emission spectra are measured against the nucleus and the
southwestern radio lobe in G29.37+0.1 (not shown here), but the 
negative velocity components are more pronounced than those observed toward the eastern side.
Consequently, the distance to both components is likely to be $\geq$17.4~kpc. 
This result disagrees with that obtained by \citet{johanson09} toward the source named
G29.3667+0.1000, which according to its name which is derived from its position, corresponds to G29.37+0.1.
It is considered by the authors as a SNR candidate located at a poorly determined distance
between 5.2 and 15.8~kpc.

Even with the HI profiles in hand, ascribing distance to G29.37+0.1 is not straightforward. 
In light of the results presented here, it is evident that the HI spectra do not 
permit us to confirm or 
reject an extragalactic origin for either the whole emission from G29.37+0.1 or some of its components, and some further information is required. 
Although only a lower distance limit was derived from the absorption profiles, 
the detection of absorption at negative velocities makes it plausible that these features 
are due to all the absorbing gas in the direction of the suspected radio galaxy traced by the S-shaped morphology.

\subsubsection{The molecular gas}
\label{CO}
\begin{figure*}[!ht]
 \centering
\includegraphics[scale=0.86]{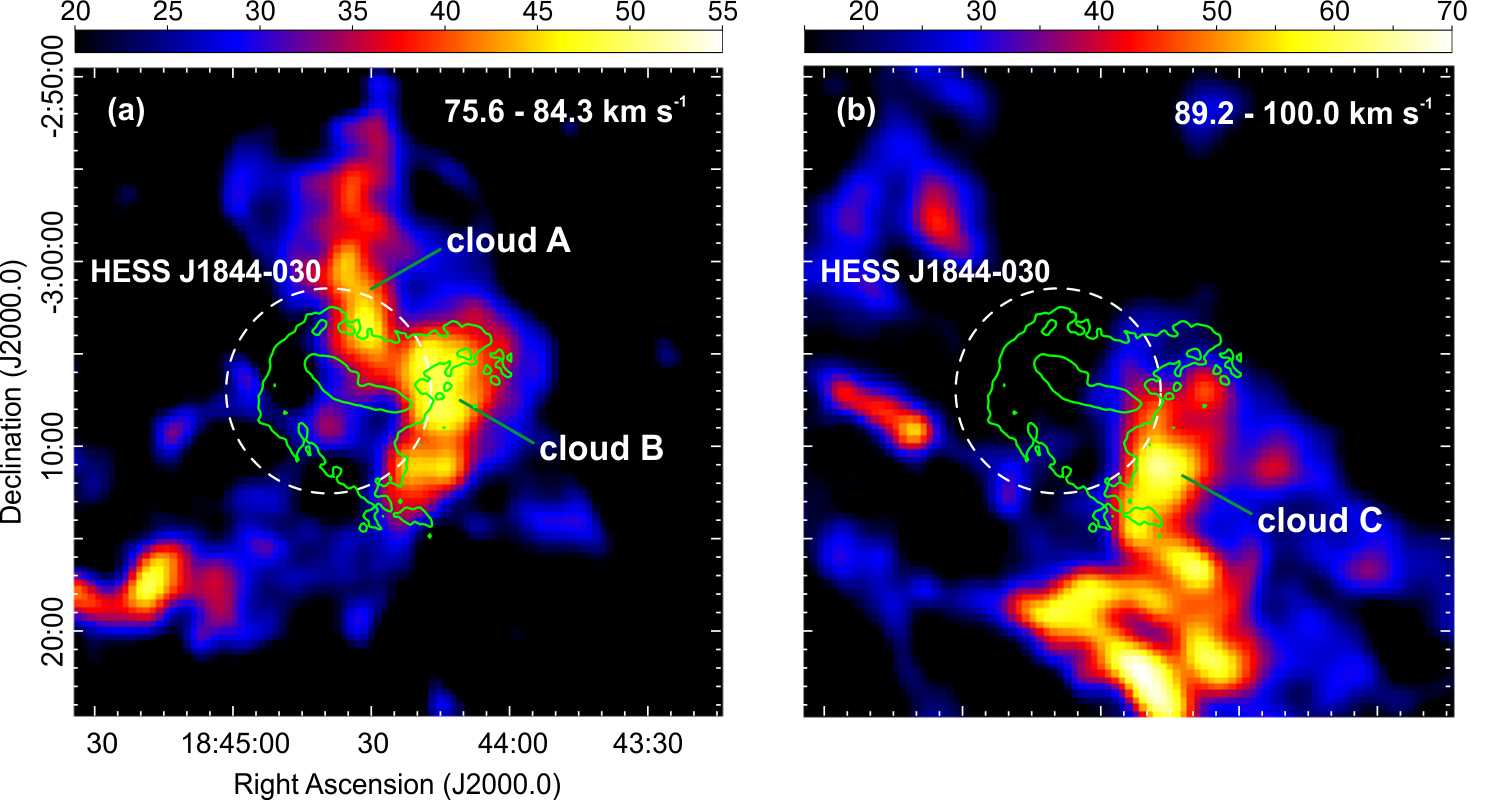}
\caption{{Emission of \CO} toward the {\G29} field, depicting the molecular clouds which show morphological agreement with the radio continuum emission from the halo component in {\G29} (green contours). Each image was obtained by integrating the {\CO} emission in a velocity range suitable to reveal the structure of the molecular clouds. 
\bf a) \rm 
{\CO} integration for clouds A and B, in the velocity range from 75.6 to 84.3~{\kms}. \bf b) \rm 
cloud C, integrated between 89.2 and 100.0~{\kms}. The white circle indicates the extension of HESS~J1844$-$030 according to the last update on this source. The scale is linear, in {K~\kms}.}
 \label{clouds}
\end{figure*}

The inspection of the molecular gas throughout the range of velocities covered by the 
GRS shows a non-uniform distribution of the $^{13}$CO lying in projection within 
the region of the G29.37+0.1/HESS~J1844$-$030 system in the velocity range from $\sim$75 to 100~km~s$^{-1}$. In particular, we identified three molecular clouds, named in what follows as clouds A,
B, and C. The morphology of these clouds may be suggestive of an association between them and 
the halo-like 
component in G29.37+0.1, for which we were not able to set a distance limit using  HI absorption spectra due to its low surface brightness. 
Up to date, a notable diversity in morphology and brightness has been identified in Galactic SNRs 
interacting with their surroundings\footnote{See http://astronomy.nju.edu.cn/~ygchen/others/bjiang/interSNR6.htm for an extensive list of Galactic SNRs interacting with molecular clouds.}  
and thus, in the alternative scenario proposed here, the halo 
morphology would represent a new case of a distorted SNR expanding into the inhomogeneities of the ISM. The detected molecular clouds are presented in Fig.~\ref{clouds},
where contours of the radio emission from the MAGPIS-1400~MHz image at 6$^{\prime\prime}$ resolution
are included for reference. Figure~\ref{clouds}a displays the {\CO} emission integrated from 75.6 to 84.3~{\kms}, the range of velocities in which clouds A and B are found, while 
Fig.~\ref{clouds}b shows the molecular emission integrated between 89.2 and 100.0~{\kms}, the range of velocities corresponding to cloud C. 

\begin{figure*}[!ht]
\centering
\includegraphics[width=0.8\textwidth]{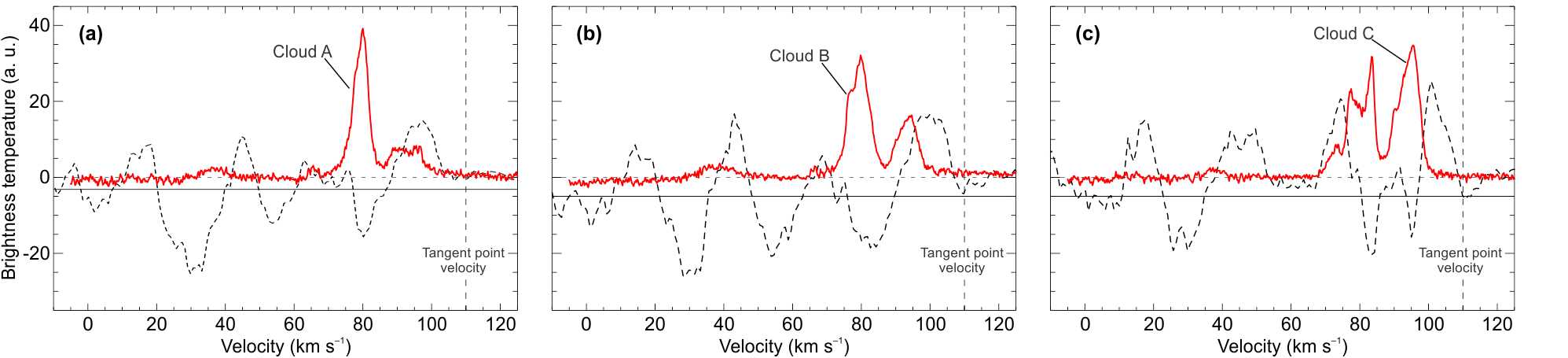}
\caption{HI absorption spectra in comparison with the {\CO} emission spectra toward the newly identified 
\bf a) \rm Cloud~A, \bf b) \rm Cloud~B, and \bf c) \rm Cloud~C, which are coincident in projection with the halo emission in {\G29}. In all cases the dashed line indicates the HI absorption spectrum obtained toward the bright clump of each cloud, while the {\CO} emission spectrum is traced by the solid line. The vertical dashed line marks the velocity of the tangent point. The horizontal, solid line indicates the rms noise in the absorption spectra of the HI.
The three cases show similar spectral features, which allow us to locate these three clouds at its near distance, 
d$_{\mathrm{clouds}}$$\sim$5-6~kpc.}
\label{CO_clouds_spectra}
\end{figure*}

While the observed alignment of the molecular emission and the halo in 
G29.37+0.1 may be coincidental, we further explore for kinematical signatures of the
presumed interacting molecular gas. 
The $^{13}$CO average spectrum for each cloud is displayed in Fig.~\ref{CO_clouds_spectra}.
They were obtained by using a $\sim$\dms{1}{\prime}{5} box centred at the brightest 
 clumps of clouds A, B, and C (the clumps are roughly 3$^{\prime}$ in size).
Along with the {\CO} emission spectra, the HI absorption profile toward each clump is shown in Fig.~\ref{CO_clouds_spectra} as dashed lines. These absorption spectra were constructed by subtracting the HI emission profile from a spectrum averaged over different regions adjacent to each molecular cloud.
For all the clouds, several HI absorption peaks are present up to the velocity of the molecular cloud, but it is noticeable that no absorption features are present at higher velocities (up to the tangent point). In order to assign a distance to the corresponding CO clouds, we used the Galactic rotation curve of \citet{Fich-89}. 
Then, the corresponding near and far distances are: 5.0/9.9~kpc for cloud A, 5.1/9.8~kpc for cloud B, and 5.8/9.0~kpc for cloud C.
To solve this kinematic distance ambiguity (KDA) and determine a unique distance to each cloud, we can observe that the clouds only marginally overlap the continuum emission from the halo, and in the weakest parts of it (Fig.~\ref{clouds}). Then, to a good approximation, we can consider that they are not superposed to any continuum source. In this scenario, the HI absorption peaks seen in the spectra of Fig.~\ref{CO_clouds_spectra} are ascribed to the absorption of the warm Galactic HI filling the interstellar medium along the line of sight by the cooler HI embedded in the molecular clouds. If the corresponding cloud were beyond the tangent point, there should not be absorption coincident with the {\CO} emission peak, since the HI gas at the same radial velocity at the near distance would be emitting too. Therefore, as this is not the case for the three clouds, 
these are in the foreground of the tangent point and we assign the near distance to all of them. Taking into account the involved uncertainties, the distances are: $5.0\pm1.8$~kpc for cloud A, $5.1\pm1.8$~kpc for cloud B, and $5.8\pm2.0$~kpc for cloud C.
Details about this procedure to solve the KDA for molecular clouds from their spectral characteristics can be found in \citet{Roman-Duval-09}.

None of the molecular clouds show direct kinematic evidence (like broadenings in the molecular spectra,
asymmetries in the profiles, etc) of interaction with the halo in G29.37+0.1. However, these spectral signatures, 
often associated with perturbations in the velocities field of the molecular gas caused by the passage of a shock front, are not conclusive. Several molecular clouds without kinematic spectral signatures are known to be 
interacting with SNRs. If this is the case, then the outer radio emitting component in G29.37+0.1 could be 
explained in terms of a supernova remnant impacting the surrounding material. 

For the sake of completeness, we  analyzed the physical properties of the molecular gas. In 
Table~\ref{fit-parameters} we summarize for each cloud the central velocity $v_c$ and the FWHM velocity, 
the peak brightness temperature, the mass, and the corresponding total proton volume density (atomic plus molecular contents).
Since the profiles are clearly non-gaussian, the listed parameters are only mean representative values.
Additionally, we followed the 
procedure presented in \citet{cas13} to calculate the H$_{2}$ column density of the identified molecular 
material and used this value, together with the corresponding distance that we derived for each cloud, to 
compute the total mass of the clouds. 
For clouds A, B, and C we assumed regions (elliptical for A, and circular for B and C) with physical sizes of approximately \dms{2}{\prime}{5} $\times$ 6{\am}, \dms{3}{\prime}{5} and 3{\am} in diameter, respectively. The uncertainties in the mass calculation are estimated to be about 30\% and arise from the velocity range in which we detect the molecular material that forms the cloud, the definition of the integration region, the estimation of the background temperature, and the assumed distances. 

\begin{table*}[ht!]
\centering
\caption{Physical properties for the $^{13}$CO clouds.}
\begin{tabular}{cccccc}                                               \hline\hline
 Cloud & $T_\mathrm{mb}$  &  $v_{c}$  & $\Delta v$  & Mass & Density  \\ 
       &  [K]             &   [km~s$^{-1}$]  &    [km~s$^{-1}$]    & [M$_{\odot}$] & [cm$^{-3}$] \\ \hline
   A   &    2.3 $\pm$ 0.4    & 80.1 $\pm$ 1.1  &  4.8 $\pm$ 0.7  &   2150$\pm$650 &   840$\pm$300    \\
   B   &    1.7 $\pm$ 0.2    & 81.5 $\pm$ 1.1  &  8.1 $\pm$ 1.7  &   1700$\pm$500 &   910$\pm$350 \\
   C   &    1.7 $\pm$ 0.3    & 94.2 $\pm$ 1.1  &  8.8 $\pm$ 1.3  &   2100$\pm$600 &  1200$\pm$400 \\\hline
\label{fit-parameters}
\end{tabular}
\end{table*}

\subsection{Plausible origins for the $\gamma$ rays from HESS~J1844$-$030}
In this section we use our findings in order to address the problem of the origin of the $\gamma$-ray emission presumably associated with {\G29}. 
The extension of the $\gamma$-ray excess in HESS~J1844$-$030 is similar to 
the point-spread function of the H.E.S.S. telescope. 
This fact favours an extragalactic origin for the production of the TeV emission  
and is consistent with our interpretation of the radio synchrotron emission from G29.37+0.1 as a new radio galaxy. To date, $\gamma$ rays in extragalactic sources have been attributed both to the leptonic process 
in which relativistic electrons in the lobes of radio galaxies are IC scattered-off by lower-energy 
ambient photons (CMB or extragalactic background light photons)  or  hadronic emission from cosmic rays interacting with the lobes \citep{mckin15}. There are some indications that 
the detected $\gamma$ rays are predominantly originated in the radio lobes rather than the 
central core region (e.g. Centaurus~A, \citealt{abdo10}). In our case 
the available TeV data do not allow distinction between
$\gamma$-ray contributions from the core and the lobes. 
In a Galactic context for HESS~J1844$-$030, in which the X-ray emitting region could be a pulsar wind nebula, 
a leptonic origin for the TeV emission via the IC process of accelerated electrons (perhaps from the 
source PS1) might be a plausible explanation.

Recently, in the region of HESS~J1843$-$033 the High Altitude Water Cherenkov (HAWC) observatory 
has detected, with a post-trial significance of 4.7$\sigma$ , the extended $\gamma$-ray source 
1HWC~J1844$-$031C \citep{abe16}. This emission, whose nature was not identified, 
is centred at RA$\sim$18$^{\mathrm{h}}$44$^{\mathrm{m}}$, 
Dec$\sim$$-$\dms{3}{\circ}{6} (RA=\dms{281}{\circ}{0}$\pm$\dms{0}{\circ}{2},
Dec=$-$\dms{3}{\circ}{1}) and separated by \dms{0}{\circ}{32} from the HESS source. Although
an association with HESS~J1843$-$033 was suggested, the morphology of the HAWC source extended
toward the $\gamma$-ray pulsar wind nebula HESS~J1846$-$029 \citep{dja08} makes this possibility
questionable.

\section{Concluding remarks}
In this work, we utilized new continuum observations from GMRT at 610~MHz together with a MAGPIS
image at 1400~MHz to conduct the first detailed study of the radio source G29.37+0.1.
In order to obtain a coherent picture of this
object, we also analyzed archival X-ray data collected by \it XMM-Newton \rm and \it Chandra\rm, 
and investigated the molecular and atomic gas in the direction of the source. Our study revealed that:

(1) Radio morphology and spectral properties: \rm At radio wavelengths, G29.37+0.1 
consists of lobes and jets
oriented symmetrically around a bright central nuclear region, 
which is reminiscent of a radio galaxy morphology. 
Furthermore, the new radio image at 610~MHz and that at 1400~MHz from MAGPIS for the source clearly reveal an extended region of weak emission, namely the halo, 
that surrounds the bright components in G29.37+0.1. 
The astrophysical connection of this structure with the lobes, jets, and the nucleus is not 
straightforward; its possible interaction with molecular material favours a Galactic origin for this emission. 
The computed global radio continuum spectrum of the S-feature in G29.37+0.1 
($\alpha$$\sim$0.6) is unambiguously produced by a synchrotron source. Additionally,
the spectral distribution over the source ($\alpha$$\sim$0.3-0.7) is compatible with values 
observed in radio galaxies. 

(2) Physical properties of the radio continuum emission: Considering an extragalactic origin for G29.37+0.1, the linear size of this source is $\sim$370~kpc for a 
redshift z$\sim$0.05 estimated using the K-z relation on the optical and infrared counterparts to the candidate radio galaxy nucleus. 
At the mentioned redshift, the estimated 610~MHz luminosity is $\sim$10$^{25}$~W~Hz$^{-1}$, 
entirely in concordance with typical luminosities of radio galaxies and thus providing 
further hints on the extragalactic nature of the source. 
The uncertainty in these calculations is dominated by our redshift determination. 
For a range 0.01<$z$<1 where most of the radio galaxies are detected, the luminosity at 610~MHz 
of G29.31+0.1 varies from $\sim$10$^{23.5}$ to $\sim$10$^{27.5}$~W~Hz$^{-1}$. 
Using simple energetic considerations 
in a revised equipartition formalism for a plasma dominated by protons (K$_{0}$=100), 
the resulting magnetic field of the source varies between
1.20 and 0.32~nT 
for the quoted redshift range.

(3) X-ray emission properties: The X rays detected in the direction 
of G29.37+0.1 are suggestive of an association with the radio lobe emission in the northeastern part of the source. Based on the X-ray spectrum alone, a single non-thermal model modified by Galactic absorption proves adequate to explain the observed diffuse X-ray emission. 
In the case of an extragalactic origin for the X rays, the magnetic field strength is  
significantly lower than those typically measured in the lobes of radio galaxies, upon the 
assumption that all of the X-ray flux is originated by IC scattering of CMB photons,
and it is also far from the field strength obtained assuming equipartition condition in the northeastern lobe of G29.37+0.1. 
A Galactic PWN is a more plausible origin for the X-ray emission and the radio halo would 
represent the shell of a SNR. The source PS1 would be the most likely pulsar candidate powering the proposed nebula. Clearly, explanations in this area deserve further detailed observations and theoretical considerations. 

(4) Surrounding environment: On the basis of HI absorption spectra against the S-shaped 
morphology, we determined a kinematic distance greater than 17.4~kpc for this component 
in G29.37+0.1.
In addition, we have identified 
a complex of strong molecular emission which lies in projection around G29.37+0.1.
The molecular material seems to match (in projection) the faint outer emission 
around the radio source, which is suggestive of an interaction.
Mass and density estimates derived from CO and HI gas components give in each cloud averaged values
of $\sim$2000~M$_{\odot}$ and 1000~cm$^{-3}$. 

(5) The shape of the TeV excess at the position of the source G29.37+0.1 is compatible with that
expected from a point source given the HESS point-spread function. This fact,
together with our interpretation of the observed radio emission as being produced in a new radio galaxy, 
suggest that HESS~J1844$-$030 could  
represent a new case of an extragalactic $\gamma$-ray emitting source.

Consequently, we propose that G29.37+0.1 is the superposition of a 
radio galaxy (the S-like feature) and a composite SNR with a shell (the radio halo component), a pulsar powered component (the diffuse X-ray emission), and perhaps even the neutron star (the source PS1). 
Our results highlight the effectiveness of high-resolution and high-sensitivity low radio frequency 
observations in resolving down to a few arcsec both morphological and spectral features across all the 
potential counterparts to the TeV $\gamma$-ray  emission. 
Radio galaxies are, of course, common features on wide-field, low frequency radio images. 
However, for lines of sight through the inner Galaxy, they are a source of confusion with SNRs 
since their spectra are similar. Our observations are important for identifying one such 
confusing source for future observers. Conversely, this newly identified radio galaxy now serves 
as a good background target, for example for measuring thermal ISM continuum absorption at even 
lower frequencies. Finally, we point out that next-generation ground-based $\gamma$-ray instruments as the Cherenkov Telescope Array, will
have the spatial resolution and sensitivity required to distinguish between the spatial contributions of $\gamma$ rays from the different components
detected at radio frequencies.

\bibliographystyle{aa}
\bibliography{aa30093-16}

\begin{acknowledgements}
The authors wish to thank the referee for his/her very useful comments which helped to improve 
the quality of the paper. 
We thank the staff of the GMRT who have made these observations possible. GMRT is run by the National Centre for Radio Astrophysics of the Tata 
Institute of Fundamental Research. 
G. Castelletti, E. Giacani and A. Petriella are members of the Carrera del Investigador Cient\'ifico of CONICET, Argentina. L. Supan is a PhD Fellow of CONICET, Argentina.
This research is supported by grants from the ANPCyT (Argentina) PICP~902/2013, the CONICET (Argentina) PIP~736/2011, and 
UBACyT 20020150100098BA.
The GRS is a joint project of Boston University and Five College Radio Astronomy Observatory, funded by the National Science Foundation under grants AST-9800334, AST-0098562, AST-0100793, AST-0228993, and AST-0507657.
This research has made use of VGPS data survey of the National Radio Astronomy Observatory, which is a facility of the National Science Foundation operated under cooperative agreement by Associated Universities, Inc. 
\end{acknowledgements}

\end{document}